\documentclass[aps,prd,groupedaddress,showpacs,twocolumn]{revtex4}
\usepackage{caption}
\usepackage{multirow}
\usepackage[parfill]{parskip}    
\usepackage{graphicx}
\usepackage{amssymb}
\usepackage{epstopdf}
\usepackage{color}
\usepackage{float}
\usepackage{ulem}
\usepackage{graphicx}
\usepackage{pdfpages}
\usepackage[colorlinks=true,citecolor=blue,urlcolor=magenta,breaklinks]{hyperref}

\begin{document}

\title{Geodesic motion around hairy black holes.}

\author{A. Ramos}
\affiliation{School of Physical Sciences \& Nanotechnology, Yachay Tech University, 100119
Urcuqu\'{i}, Ecuador}

\author{C. Arias}
\affiliation{School of Physical Sciences \& Nanotechnology, Yachay Tech University, 100119
Urcuqu\'{i}, Ecuador}

\author{R. Avalos }
\affiliation{Departamento de F\'isica, Colegio de Ciencias e Ingenier\'ia, Universidad San Francisco de Quito,  Quito, Ecuador\\}

\author{E. Contreras }
\email{econtreras@usfq.edu.ec}
\affiliation{Departamento de F\'isica, Colegio de Ciencias e Ingenier\'ia, Universidad San Francisco de Quito,  Quito, Ecuador\\}

\begin{abstract}
In a recent paper (Phys. Dark Univ. {\bf 31}, 100744 (2021)) it has been obtained new static black hole solutions with primary hairs by 
the Gravitational Decoupling.
In this work we either study the geodesic motion of massive and massless particles around those solutions and restrict
the values of the primary hairs by observational data. In particular, we obtain the effective potential, the innermost stable circular orbits, the marginally bounded orbit, and the periastron advance for time--like geodesics. In order to restrict the values taken by the primary hairs we explore their relationship with the rotation parameter of the Kerr black hole giving the same innermost stable circular orbit radius and give the numerical values for the supermassive black holes at Ark 564 and NGC 1365.
The photon sphere and the impact parameter associated to null geodesics are also discussed. 
\end{abstract}

\maketitle

\section{Introduction}\label{intro}
Through the developments of general relativity black holes (BH) solutions have been a subject of study, discussions and analysis. From being considered as simply mathematical constructions lacking for physical reality to be one of the central topics in recent researches, BH are undoubtedly one of the most known and intriguing objects in literature. Indeed, the most recent Nobel Prize in Physics has been given not only for theoretical developments on BH's \cite{penrose1965} but for the studies of orbits followed by massive objects (stars) around the presumably super massive BH in the center of the Milky Way, Sagittarius A* \cite{g1,g2,g3,g4}. In this sense, it is clear that the study of the motion around BH's by solving the geodesic equations of test particles it is important not only to show that they indeed are real objects but with the aim to understand some of their main features given that study of geodesics exposes geometric properties of space--time. More precisely, the geodesic equations can be applied to calculate observable quantities like the shadow of a black hole, the periastron shift of a bound orbit and many other quantities which allows to characterize the BH geometry \cite{synge,luminet,bardeen,carlos1,carlos2,Bambi:2008jg,Bambi:2010hf,study1,study2,Moffat,quint2,study3,Schee:2016yzb,study4,study5,Schee:2017hof,study6,bobir2017,kumar2018,ovgun2018,Konoplya:2019sns,sudipta2019a,shakih2019b,contreras2019rot,sabir19,sunnya,sunnyb,Konoplya:2019xmn,Konoplya:2019fpy,Allahyari:2019jqz,Tinchev:2019qwt,Cunha:2019ikd,Ovgun:2019jdo,Konoplya:2019goy,Hensh:2019ipu,Stuchlik:2019uvf,Contreras:2019cmf,Fathi:2019jid,Chang:2020lmg,Badia:2020pnh,Li:2020drn,Li:2019lsm,Khodadi:2020jij,Ovgun:2020gjz,Liu:2020ola,Vagnozzi:2020quf,Ghosh:2020ece,stuchlikprd}.
\\

The non--hair conjecture states that, independent of the manner in which a BH is formed, it can be only characterized by three parameters namely, the mass, charge and angular momentum \cite{hawking1,hawking2}. However, it has been demonstrated that 
under certain circumstances the existence of a BH characterized with more than these three parameters is a possibility \cite{h1,h2,h3,h4,h5,h6,h7,h8,h9,h10,h11}. In this sense, taken hairy BH as real possibility it could be interesting to explore how the geodesic motion of particles around it deviates from their behaviour when the hairs are ignored.\\

Among all the possible hairy BH in literature we could consider, in this paper we concentrate on one recent solution reported in \cite{Ovalle:2020kpd}. This hairy BH were obtained by the now well--known Gravitational Decoupling (GD) \cite{ovalle2017} trough the Minimal Geometric Deformation (MGD) \cite{ovalle15,ovalle15a}
(for implementation in $3+1$ and $2+1$ dimensional spacetimes see 
\cite{rocha2017a,rocha2017b,casadio2017a,ovalle2018,ovalle2018bis,estrada2018,ovalle2018a,lasheras2018,gabbanelli2018,sharif2018,sharif2018a,sharif2108b,fernandez2018,fernandez2018b,contreras2018,estrada,contreras2018a,morales,tello18,rincon2018,ovalleplb,contreras2018c,contreras2019,contreras2019a,tello2019,contrerasextended,tello2019a,lh2019,estrada2019,gabbanelli2019,ovalle2019a,sudipta2019,victorNS,linares2019,leon2019,casadioyo,singh19,maurya19a,sharif19a,tello2019c,abellan20,Sharif:2020pze,Sharif:2020rlt,sharif20a,tello20,maurya20b,rincon20a,sharif20b,maurya20c,jorgeLibro,Abellan:2020dze,zubair} and references there in)  in its extended version \cite{ovalleplb}. As it is reported in \cite{Ovalle:2020kpd}, this black holes were obtained by demanding they should fulfill the strong (SEC) or the dominant (DEC) between the horizon of the BH to infinity. As demonstrated in \cite{Ovalle:2020kpd}, all the new hairy BH's solutions correspond to deformations of the Schwarzschild vacuum. It is our main goal here to explore how the 
presence of such primary hairs which deform the Schwarzschild geometry affects the geodesics of massive and massless test particles around. In particular we study
the innermost stable circular orbits (ISCO), the marginally bound orbit (MBO) for massive test particles. Additionally we obtain the periastron advance by numerical computations and  analyze how much the orbits are affected in comparison to the Schwarzschild case. Besides, we explore the values that should be taken by the primary hairs in order to mimic the same effect that Kerr solution has on test particles at the ISCO radius. This analysis allows to restrict the values of such a a primary hairs using observed data. Finally,
we also consider the behaviour of null geodesics to see how the hairs affect the photon sphere.\\

This work is organized as follows. In the next section we review the main aspects on geodesic motion around a central object. Next, in section \ref{hairy} we introduce the hairy BH's reported in \cite{Ovalle:2020kpd}. 
Section \ref{timelike} is devoted to the analysis of the impact parameters, ISCO and MBO for each of the BH models introduced. In section  
\ref{bo} the bounded orbits are studied and next, in section \ref{mimic}, we obtain the numerical values of the primary hairs that mimic the spin parameter of the Kerr solution. In section \ref{null} an analysis 
on null geodesics is performed and
we conclude our work in the last section.

\section{Geodesic equations }

In this section we review the basic concepts related to the geodesic motion around black holes. The geodesic motion of test particles in a spherically symmetric space-time with metric
\begin{eqnarray}
ds^{2}=f dt^{2} -f^{-1} dr^{2}-r^{2}d\Omega^{2},
\end{eqnarray}
is described by the geodesic equations
\begin{eqnarray}
\dot{r}^{2}&=&E^{2}-\frac{f}{r^{2}}\left(\mathcal{Q}+L^{2}+\epsilon r^{2}\right)\label{rdot}\\
\dot{\theta}^{2}&=&\frac{1}{r^{4}}\left(\mathcal{Q}-L^{2}\cot^{2}\theta\right)\\
\dot{\phi}&=&\frac{L}{r^{2}}\csc^{2}\theta\\
\dot{t}&=&\frac{E}{f},
\end{eqnarray}
where $\epsilon=0$ for null geodesics, $\epsilon=1$ for timelike geodesics, $\mathcal{Q}$ is the Carter constant, and $L$ and $E$ are the specific angular momentum and energy, respectively. The orbits of particles can be described by the effective potential  
\begin{eqnarray}\label{vefe}
V_{eff}^{2}=\frac{f}{r^{2}}\left(\mathcal{Q}+L^{2}+\epsilon r^{2}\right),
\end{eqnarray}
which can read from Eq. \ref{rdot}.
In this work we shall concentrate on equatorial motion, namely $\theta=\pi/2$ and $\dot{\theta}=0$ which leads to $\mathcal{Q}=0$, from where (\ref{vefe}) reduces to
\begin{equation} \label{eq:veff}
    V_{eff}^2 = \frac{f}{r^2} \left( \epsilon + \frac{L^2}{r^2}\right).  
\end{equation}

The orbits can be classified in terms of the real roots of the the inequality
\begin{eqnarray}
V_{eff}\le E.
\end{eqnarray}
When $E>V_{eff}$ the particle comes from infinity and moves directly to the origin and this orbit is called the terminating escape orbit. In the case that $E=V_{eff}$ has one real root, the particle can either 
move from a finite radius $r$ direct to the origin or move on an escape orbit. When $E=V_{eff}$ leads to two real roots, the particle can describe a bounded orbit (as planetary motion) \cite{chakra,Chandrasekhar:1985kt,Bambi:2017khi}.\\

To obtain the impact parameters associated with the bounded orbits of the particles, we apply the standard conditions $\dot{r} = 0$, $\ddot{r}=0$, and set $V_{eff}'= 0$, where the prime $'$ denotes derivative with respect to the radial coordinate. The bounded orbits can be either stable or unstable, where the stability is determined by $V_{eff}''$. On one hand, if $V_{eff}'' > 0$ the orbit corresponds to a maximum point in the effective potential, where a small perturbation would destabilize the orbit and therefore the bounded orbit is unstable. In contrast, if $V_{eff}''<0$ small perturbation would lead to small oscillations around the orbit; therefore it corresponds to a stable orbit. We focus on the study of two important bounded orbits, namely ISCO and MBO. The ISCO, which is the smallest stable orbit around a massive object, corresponds to the inflection point of the effective potential by setting $V''_{eff}=0$, which provides information associated with the accretion disk of BH. The MBO corresponds to the critical bound orbit with energy $E=1$ that separates the bounded orbits ($E<1$) from the unbounded orbits ($E>1$), which plays a relevant role in the description of the dynamics of star clusters around supermassive black holes \cite{Clifford:2012}. 

\section{hairy BH solutions}\label{hairy}
In this section we briefly review the hairy BH solutions obtained in reference \cite{Ovalle:2020kpd}. It is worth mentioning that all the models in which we will base our geodesic analysis were previously obtained by gravitational decoupling (GD) which has become in a powerful tool in the study of BH \cite{ovalle2018a}, interior solutions \cite{ovalle2017}, $f(R)$ \cite{sharif19a}, $f(R,T)$ \cite{tello2019c} and $f(\mathcal{G})$\cite{Sharif:2020rlt} gravity, Brans--Dicke \cite{sharif20a} and cosmology \cite{linares2019}. In the particular case \cite{Ovalle:2020kpd}, GD was used to obtain hairy BH by the extended minimal geometric deformation \cite{ovalleplb} and all the solutions satisfy either SEC or DEC outside the horizon and contain the Schwarzschild BH as a limit when the deformation parameter is turned off as we shall discuss as follows.

\subsection{Model 1}
This BH solution is characterized by a metric function of the form
\begin{eqnarray}\label{orig-model1}
f=1-\frac{2\mathcal{M}}{r}+\alpha e^{-r/(\mathcal{M}-\alpha\ell/2)},
\end{eqnarray}
where $\mathcal{M}=M+\alpha \ell/2$, $\alpha$ is the decoupling parameter which connects (\ref{orig-model1}) with
the Schwarzschild BH of mass $M$ when $\alpha=0$. At this point some comments are in order. First, note that (\ref{orig-model1}) approaches asymptotically to a Schwarzschild solution with a shifted mass. Second, the solution satisfy the SEC whenever $r\ge 2M$. Moreover, it can be shown that  SEC leads to inequalities which are invariant under transformations involving    $\alpha\ell\equiv\ell_{0}$ so this parameter acts as a ``gauge charge'' 
(see \cite{Ovalle:2020kpd} for details). Finally, the location of the horizon of (\ref{orig-model1}) satisfies
\begin{eqnarray}
\ell_{0}=r_{H}-2M+\alpha r_{H}e^{-r_{H}/M}.
\end{eqnarray}
Note that demanding $r_{H}\ge 2M$ (SEC) entails $\ell\ge 2M/e^{2}$ which in the extremal case leads to a solution with BH horizon located at $r_{H}=2M$. In this case the metric function reads
\begin{eqnarray}\label{model1}
f=1-\frac{2M}{r}+\alpha \left(e^{-r/M}-\frac{2M}{r}e^{-2}\right).
\end{eqnarray}
Note that we can base our analysis on either Eq. (\ref{orig-model1}) or Eq. (\ref{model1}). However, we shall concentrate in the latter given that for the special value taken by $\ell_{0}$ the radius of the modified solution coincides with the Schwarzschild case.

\subsection{Model 2}
This BH solution is characterized by a lapse function which reads
\begin{eqnarray}\label{orig-model2}
f=1-\frac{2\mathcal{M}}{r}+\frac{Q^{2}}{r^{2}}-\frac{\alpha (\mathcal{M}-\ell_{0}/2) e^{-r/(\mathcal{M}-\ell_{0}/2)}}{r},
\end{eqnarray}
where $\mathcal{M}=M+\ell_{0}/2$, $\ell_{0}=\alpha\ell$ and $Q$ is a constant with dimensions of a length proportional to $\alpha$ and, again, the Schwarzschild BH of mass $M$ is recovered when $\alpha\to 0$. This solution satisfy the DEC for $r\ge 2M$ and the location of the horizon radius is obtained after solving
\begin{eqnarray}\label{hc}
\ell_{0}=r_{H}-2M+\frac{Q^{2}}{r_{H}}
-\alpha M e^{-r_{H}/M}.
\end{eqnarray}
Note that, using (\ref{hc}) in
(\ref{orig-model2}) we obtain
\begin{eqnarray}\label{metric-seed-DEC}
f=1-\frac{r_{H}}{r}\left(1+\frac{Q^{2}}{r_{H}^{2}}-\frac{\alpha M}{r_{H}}e^{-r_{H}/M}\right)+\frac{Q^{2}}{r^{2}}-\frac{\alpha M}{r}e^{-r/M}
\end{eqnarray}
Note that, the fulfilling of DEC requires that $r_{H}\ge 2M$ which leads to additional restrictions to $Q$ and $\ell$, namely
\begin{eqnarray}
Q^{2}&\ge& 4\alpha(M/e)^{2}\\
\ell&\ge& M/e^{2}
\end{eqnarray}
It is straightforward to show that, when the above conditions are saturated, Eq. (\ref{metric-seed-DEC}) becomes 
\begin{eqnarray}\label{model2}
f=1-\frac{2M}{r}\left(1+\frac{\alpha}{2e^{2}}\right)+\frac{4\alpha M^{2}}{e^{2}r^{2}}-\frac{\alpha M}{r}e^{-r/M},
\end{eqnarray}
which, as in the previous case, the horizon is located at $r_{H}=2M$.
It is worth mentioning that Eq. (\ref{model2}) can be rewritten in an alternative way after defining $\mathcal{M}=M\left(1+\frac{\alpha}{2e^{2}}\right)$, namely
\begin{eqnarray}
f=1-\frac{2\mathcal{M}}{r}+\frac{Q^{2}}{r^{2}}
-\frac{\sqrt{\alpha}Q}{2r}e^{1-2\sqrt{\alpha}r/e Q},
\end{eqnarray}
which can be interpreted as a BH solution supported by a non--linear electrodynamics source. 

\subsection{Model 3}
This model is obtained by assuming 
\begin{eqnarray}\label{charge}
Q^{2}&=&\alpha\ell M(2+\alpha e^{-\alpha\ell/M}),
\end{eqnarray}
from where the horizon condition
(\ref{hc}) is fulfilled whenever $r_{H}=\alpha\ell=\ell_{0}$. Using this condition and replacing (\ref{charge}) in (\ref{metric-seed-DEC}) we have
\begin{eqnarray}
f=1-\frac{2M+\ell_{0}}{r}+
\frac{2\ell_{0}M}{r^{2}}-
\frac{\alpha M e^{-r/M}}{r^{2}}(r-\ell_{0}e^{\frac{r-\ell_{0}}{M}})
\end{eqnarray}

\subsection{Model 4}
This model is obtained by considering
\begin{eqnarray}
Q^{2}=\alpha M(2M+\alpha\ell)e^{-\frac{(2M+\alpha\ell)}{M}},
\end{eqnarray}
in (\ref{metric-seed-DEC}) which leads to 
\begin{eqnarray}
f=1-\frac{2M+\ell_{0}}{r}-\frac{\alpha M}{r^{2}}e^{-r/M}(r-(2M+\ell_{0})e^{\frac{r-2M+\ell_{0}}{M}}),
\end{eqnarray}
which corresponds to a BH solution obeying the DEC with horizon located at $r_{H}=2M+\ell_{0}$.\\

We conclude this section by highlighting that all the models under consideration 
represent BH's which are characterized by the parameters $M$, $\alpha$ and $\ell_{0}$ with $\{\alpha,\ell_{0}\}$
representing primary hairs.

\section{Timelike geodesics}\label{timelike}

In this section, we study the impact parameters, ISCO, MBO and bounded orbits for each of the BH models, presented in the previous section, and compared with the Schwarzschild BH. 

\subsection{Effective Potential}
\begin{figure*}[htb!]
\centering
\includegraphics[width=0.30\textwidth]{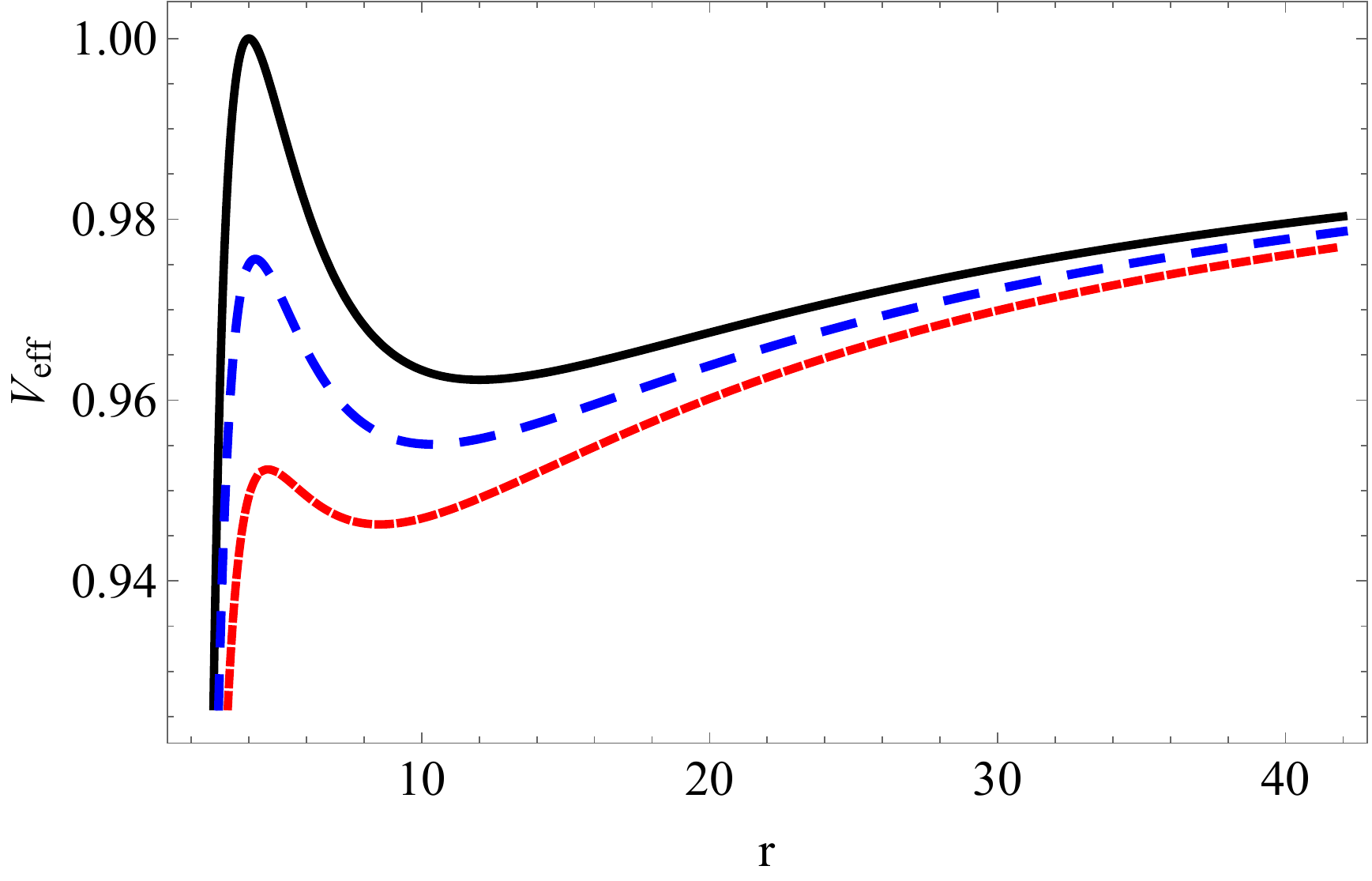}  \
\includegraphics[width=0.30\textwidth]{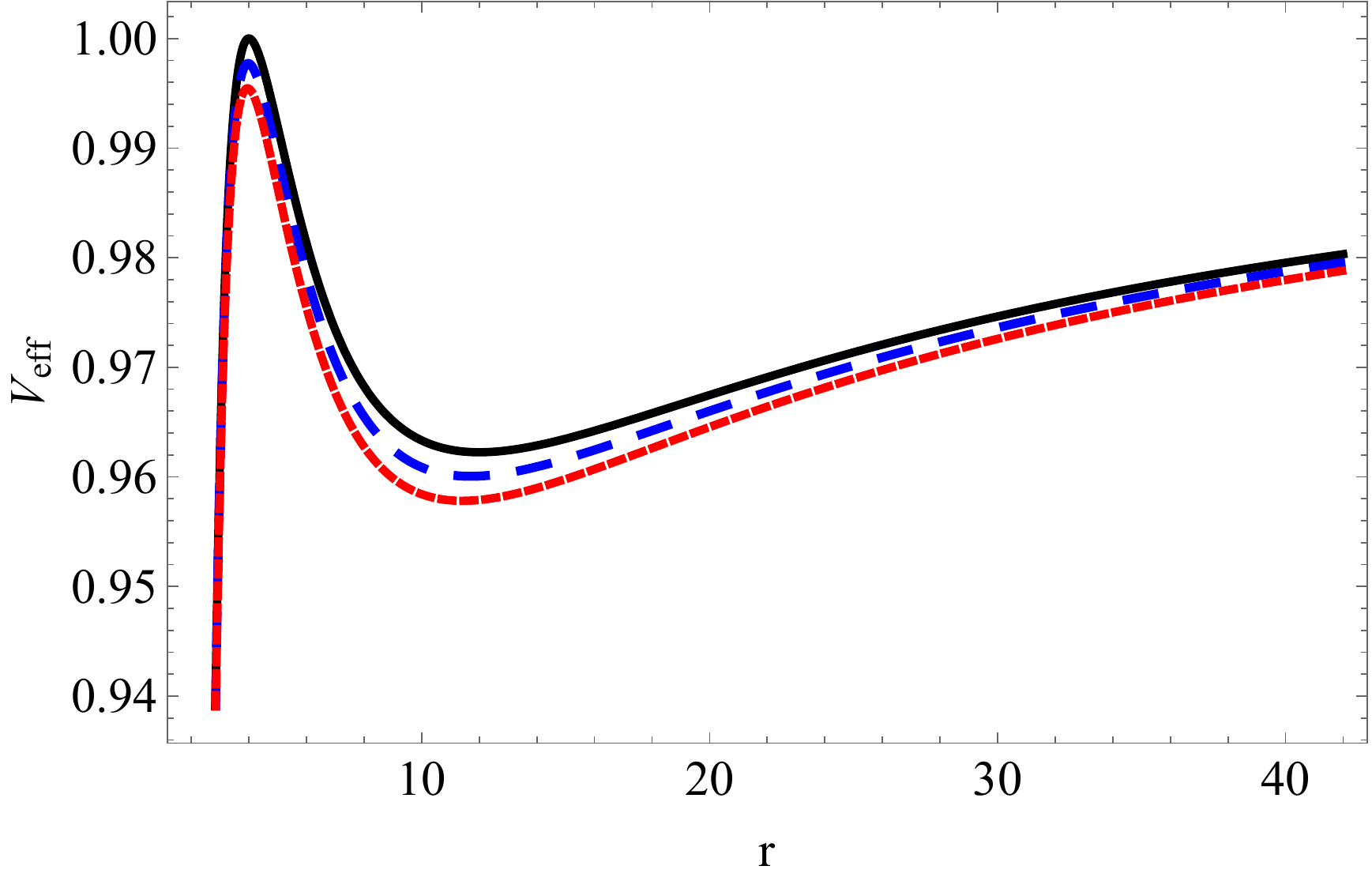}  \
\medskip

\includegraphics[width=0.30\textwidth]{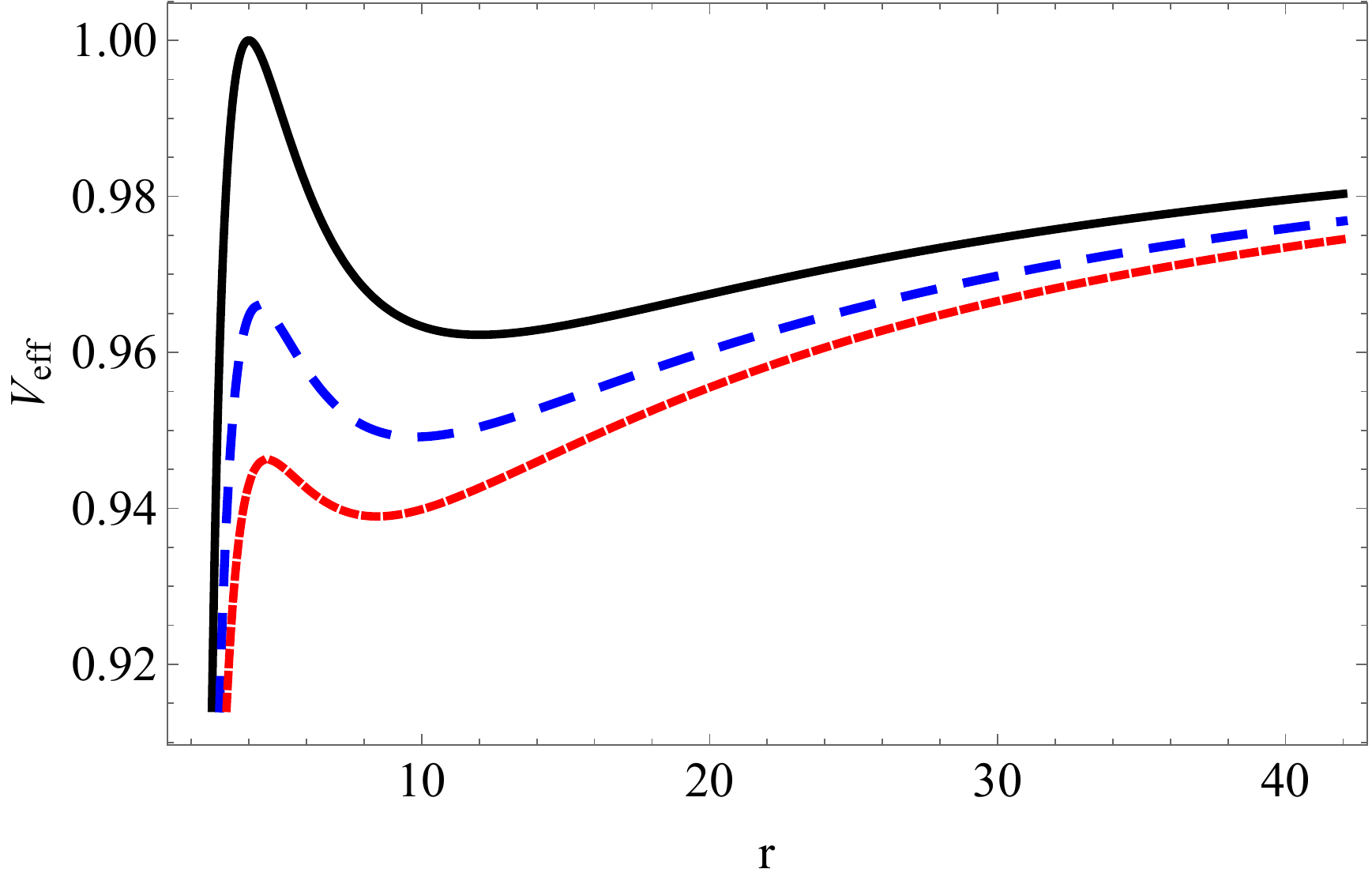}  \
\includegraphics[width=0.30\textwidth]{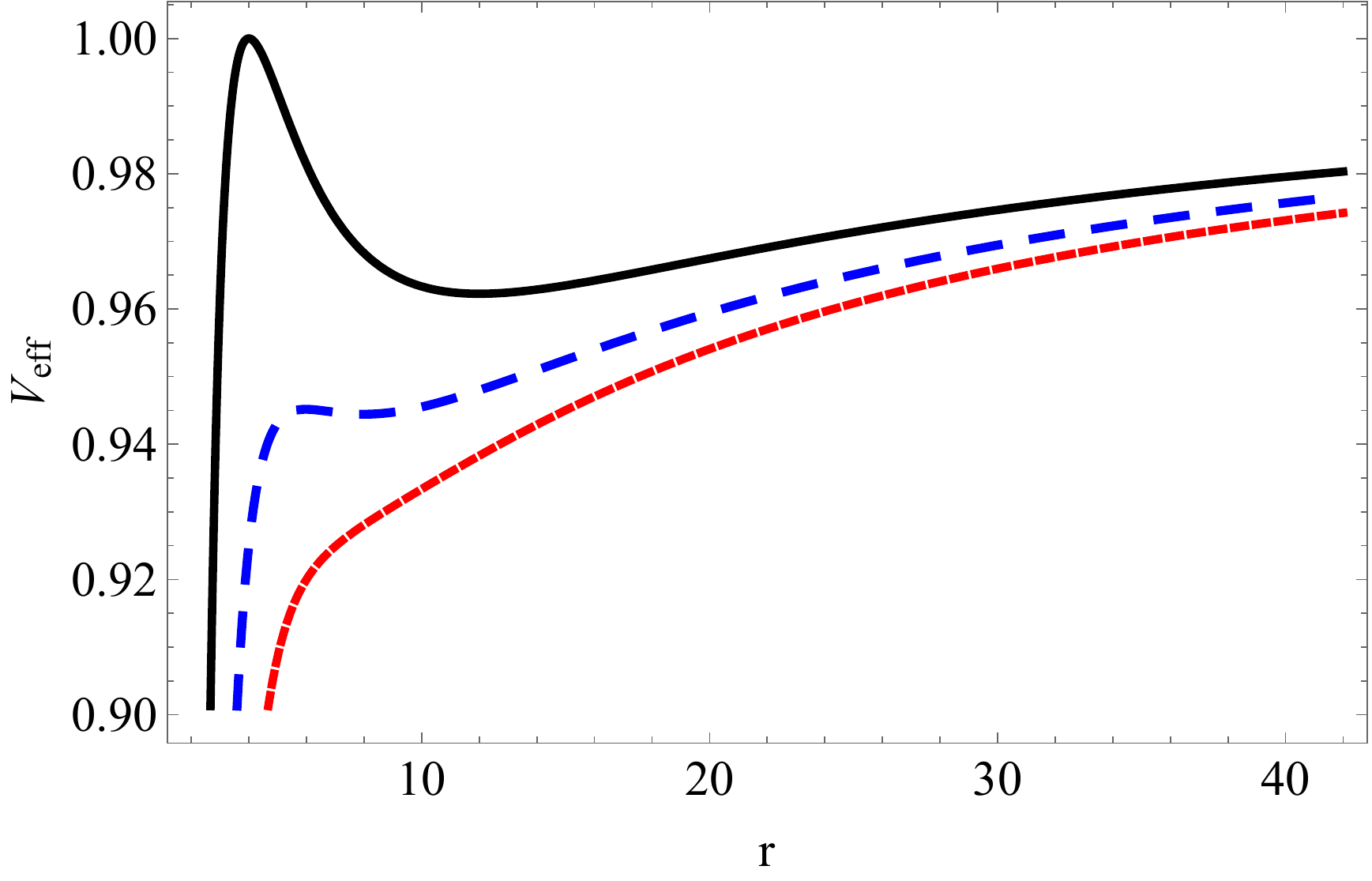}
\caption{\label{veff} 
First row. Effective potential for equatorial geodesics for $M=1$, $L= 4$ and $\alpha=0$ (black line, Schwarzschild), $\alpha=0.5$ (blue dashed line) and $\alpha=1$ (red dotted line) for model 1 (left panel) and model 2 (right panel)\\
Second row. Effective potential for equatorial geodesics for $M=1$, $L= 4$ and $\ell_{0}=0$ (black line, Schwarzschild), $\ell_{0}=0.3$ (blue dashed line) and $\ell_{0}=0.5$ (red dotted line) for model 3 (left panel)
and model 4 (right panel)}. 
\end{figure*}

The profiles of the effective potentials for the Hairy BH solutions are illustrated in  Fig. \ref{veff}
for different values of $\{\alpha,\ell_{0}\}$.
Each panel corresponds to a particular model as indicated in the legend of the figure. The black line corresponds to the effective potential associated to the Schwarzschild BH while both the red and the blue lines correspond to the hairy solution for certain values of $\alpha$ (models 1 and 2) and $\ell_{0}$ (models 3 and 4). It is noticeable how all the profiles deviate from the Schwarzschild case for $r_{H}<r<\infty$.
However, all the potentials approach asymptotically to Schwarzschild  whenever  
$r\approx r_{H}$ or $r\to\infty$. Now, in all the cases, energies associated to bounded orbits in Schwarzschild background, corresponds to unbounded orbits around the hairy BH's. More precisely, it is required  lower energies to describe bounded orbits in the four models under consideration. Additionally, the minimum of the potential is reached at lower radius
as $\alpha$ and $\ell_{0}$ grow, which means that the size of the circular orbits decreases.

\subsection{Impact parameters}
In this section we shall explore the behavior of the impact parameters for bounded geodesics per each of the hairy black holes introduced in \ref{hairy}. To this end we implement the standard conditions $\dot{r}=0$ and
$\ddot{r}=0$ in Eq. (\ref{rdot}) from where
\begin{eqnarray}
L^{2}&=&  \frac{2 f^2}{2 f-r f'}\label{eq:L2_M1}\\
E^{2}&=& \frac {2 r^2 f } {2f - r f' } - Q - r^2 \label{eq:E2_M1}
\end{eqnarray}
with $f$ given by the corresponding hairy BH.\\

\begin{figure*}[hbt!]
\centering
\includegraphics[width=0.24\textwidth]{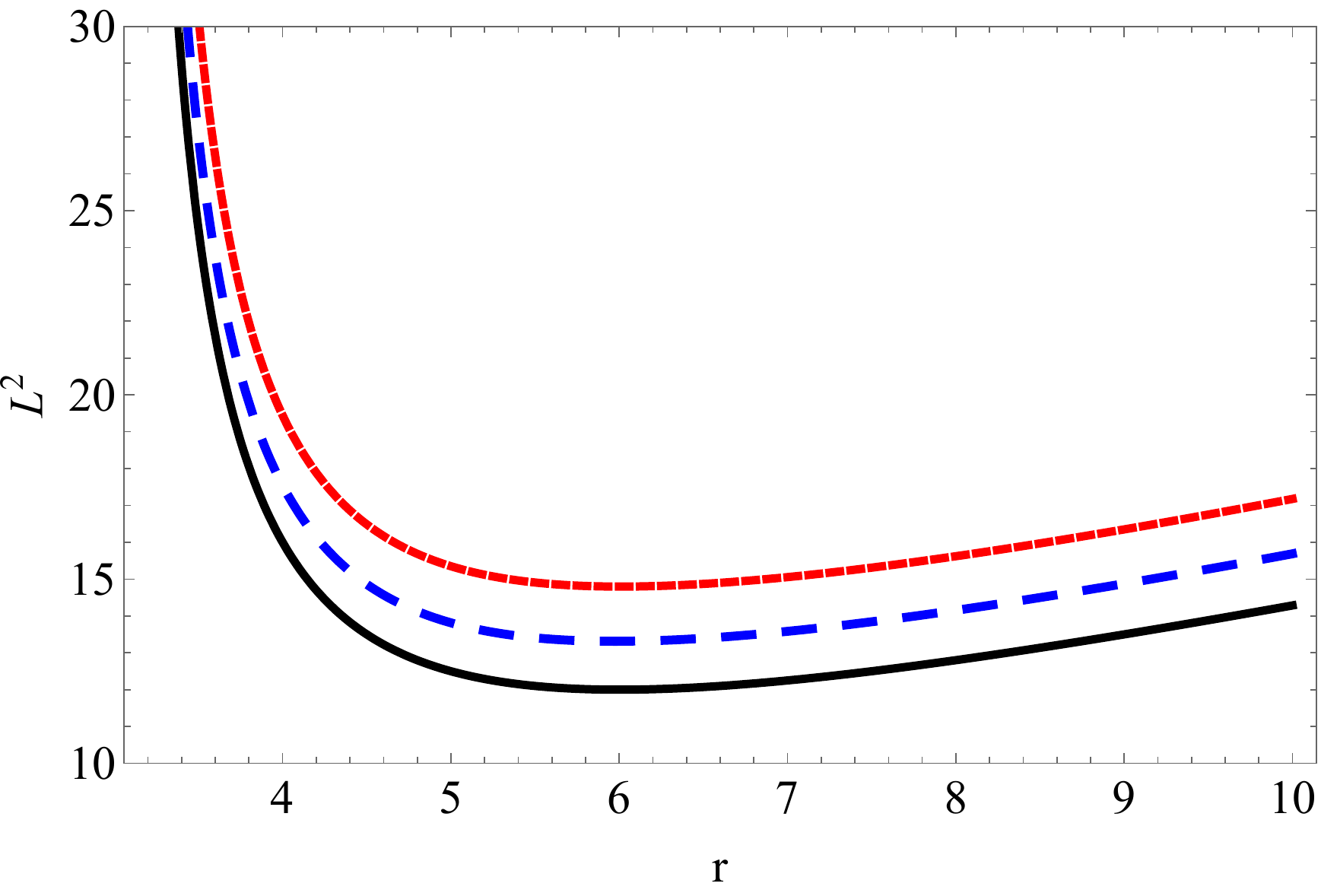}  \
\includegraphics[width=0.24\textwidth]{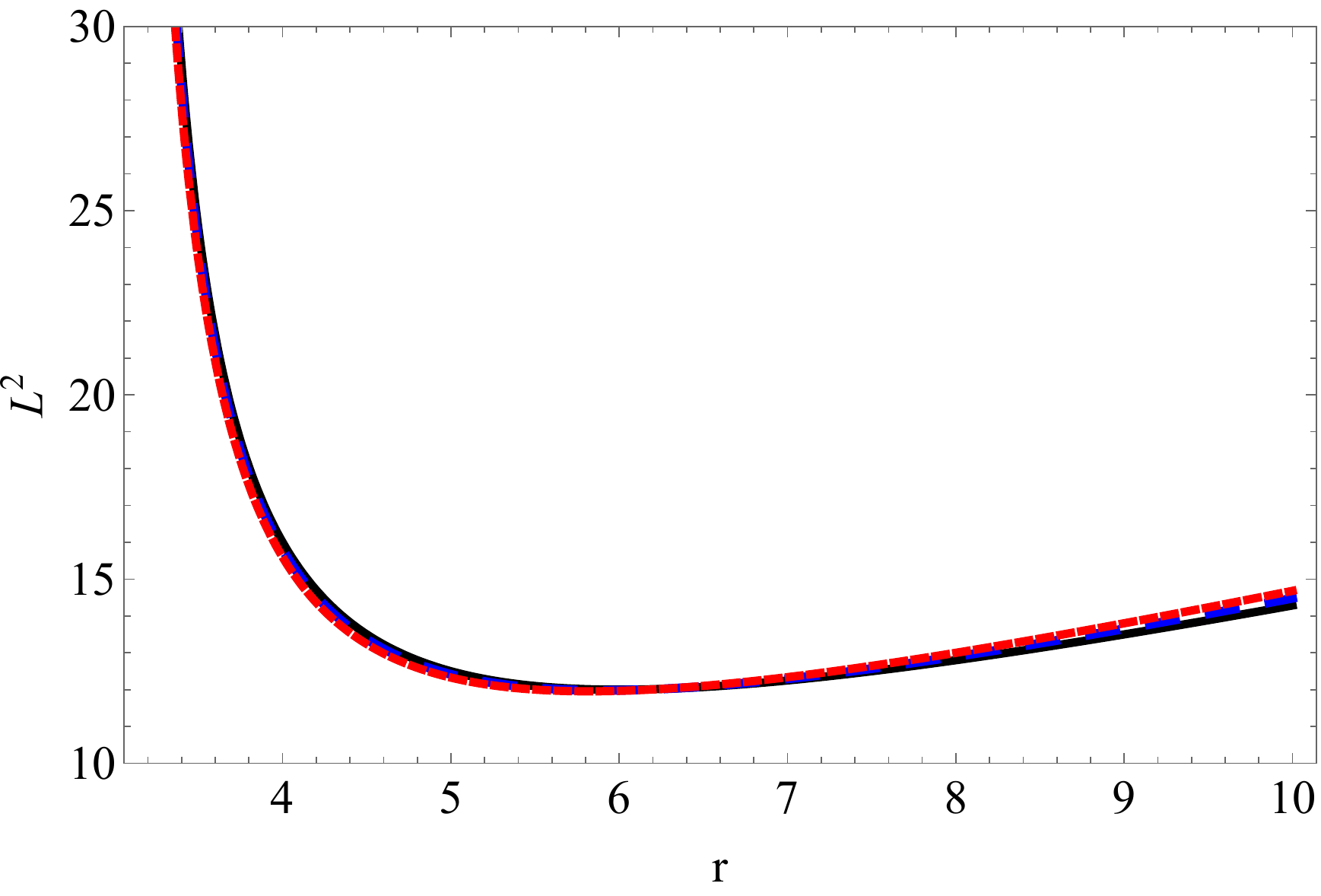}  \
\includegraphics[width=0.24\textwidth]{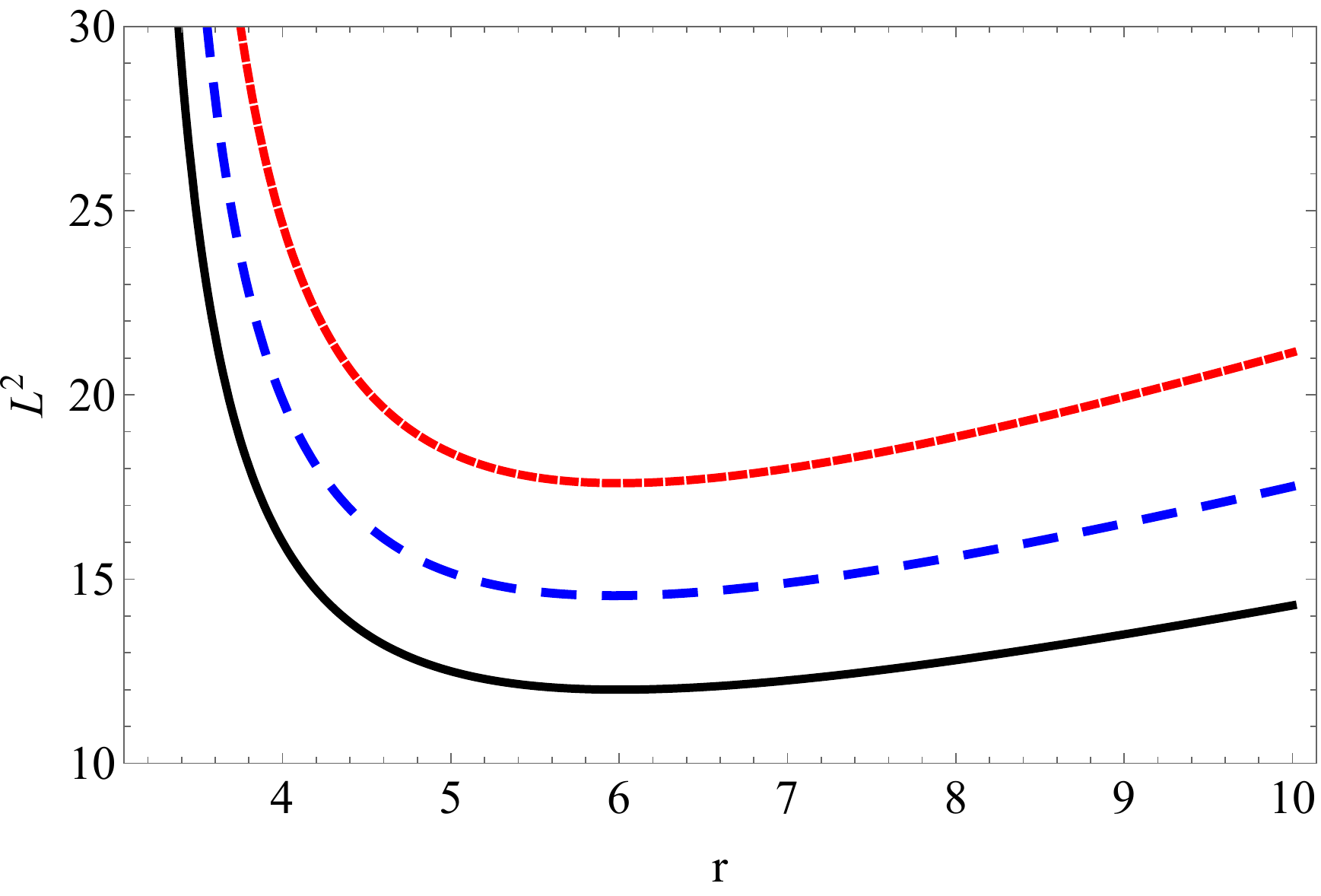}  \
\includegraphics[width=0.24\textwidth]{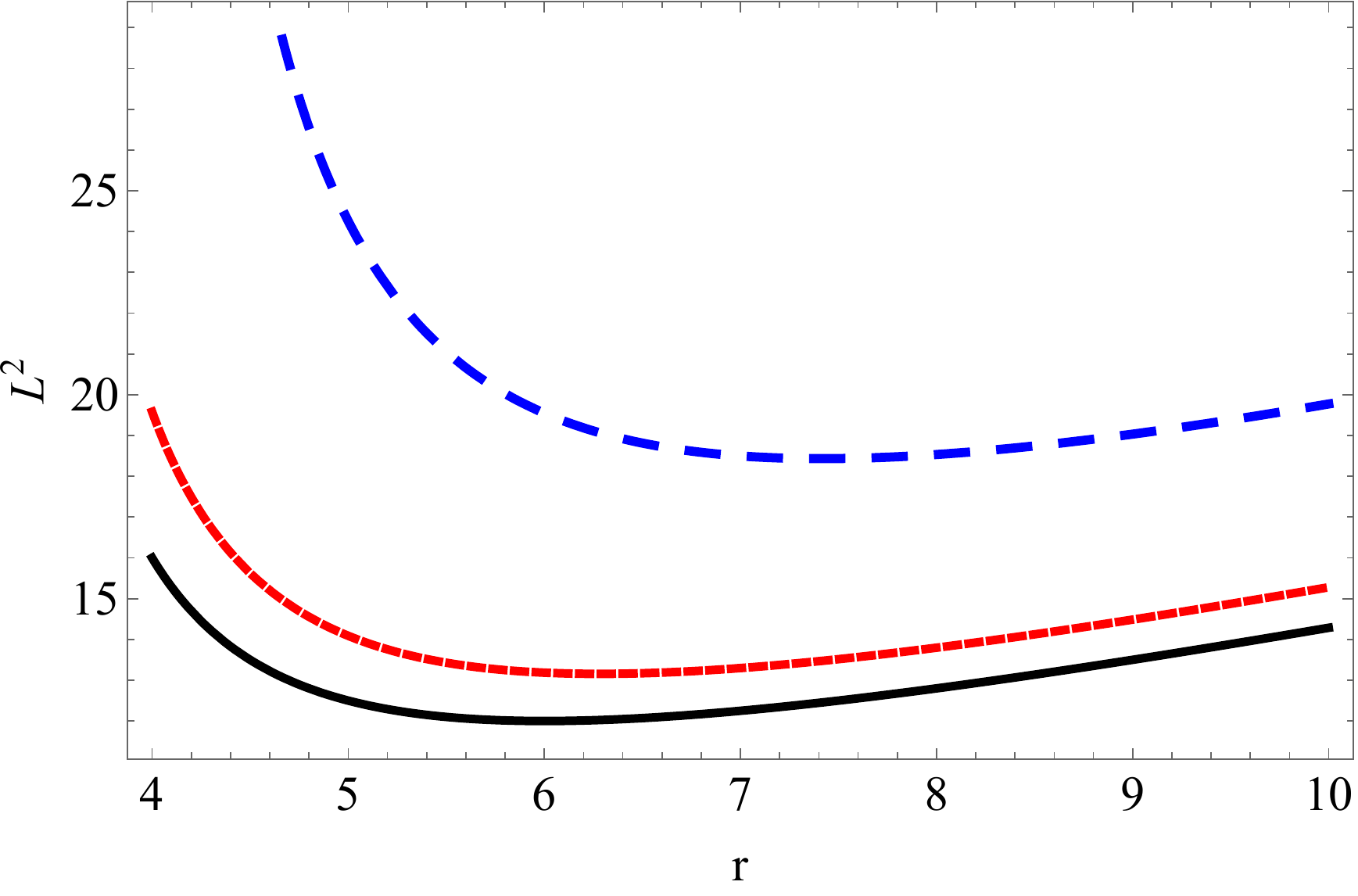}  \
\medskip

\includegraphics[width=0.24\textwidth]{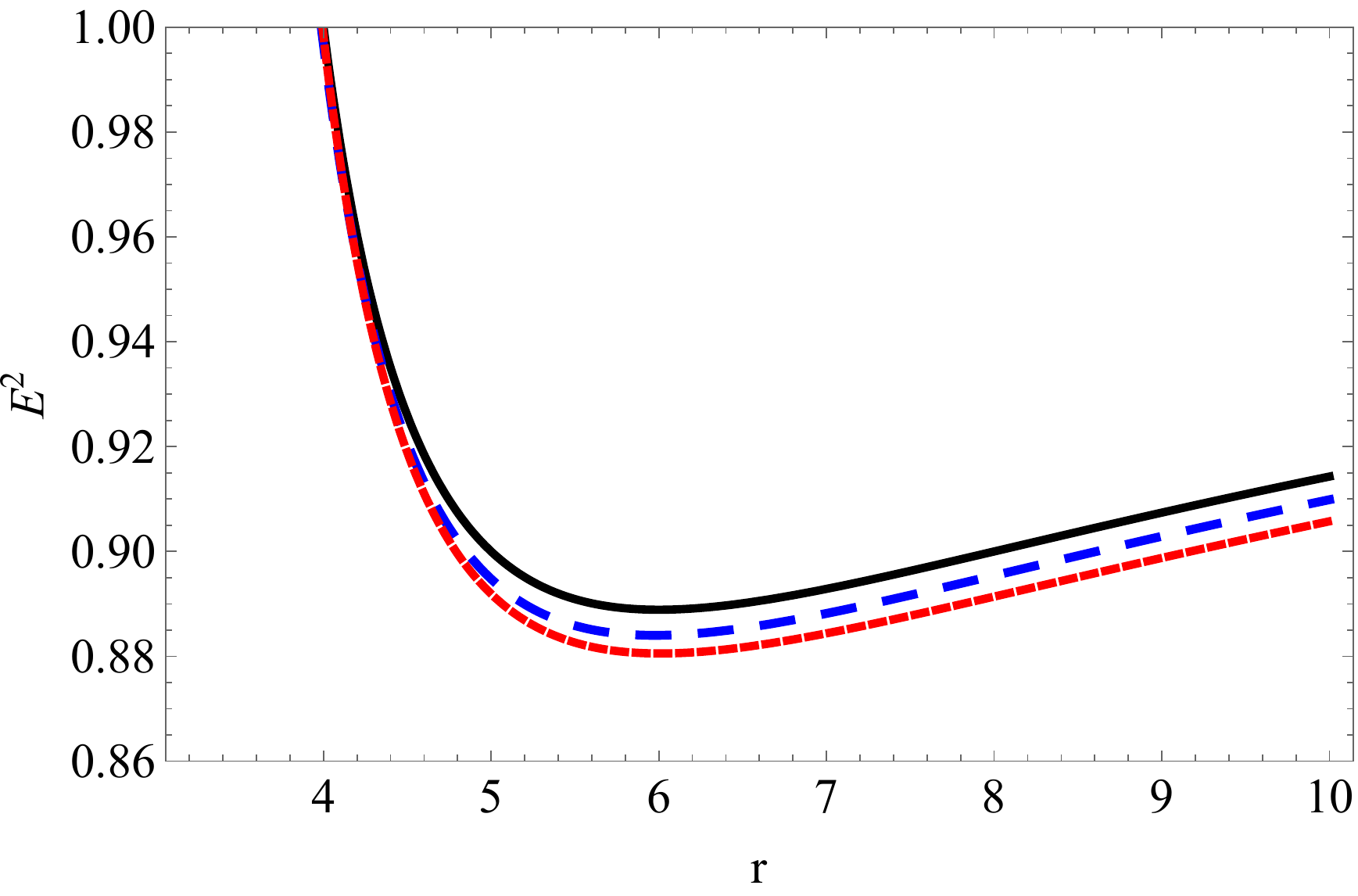}  \
\includegraphics[width=0.24\textwidth]{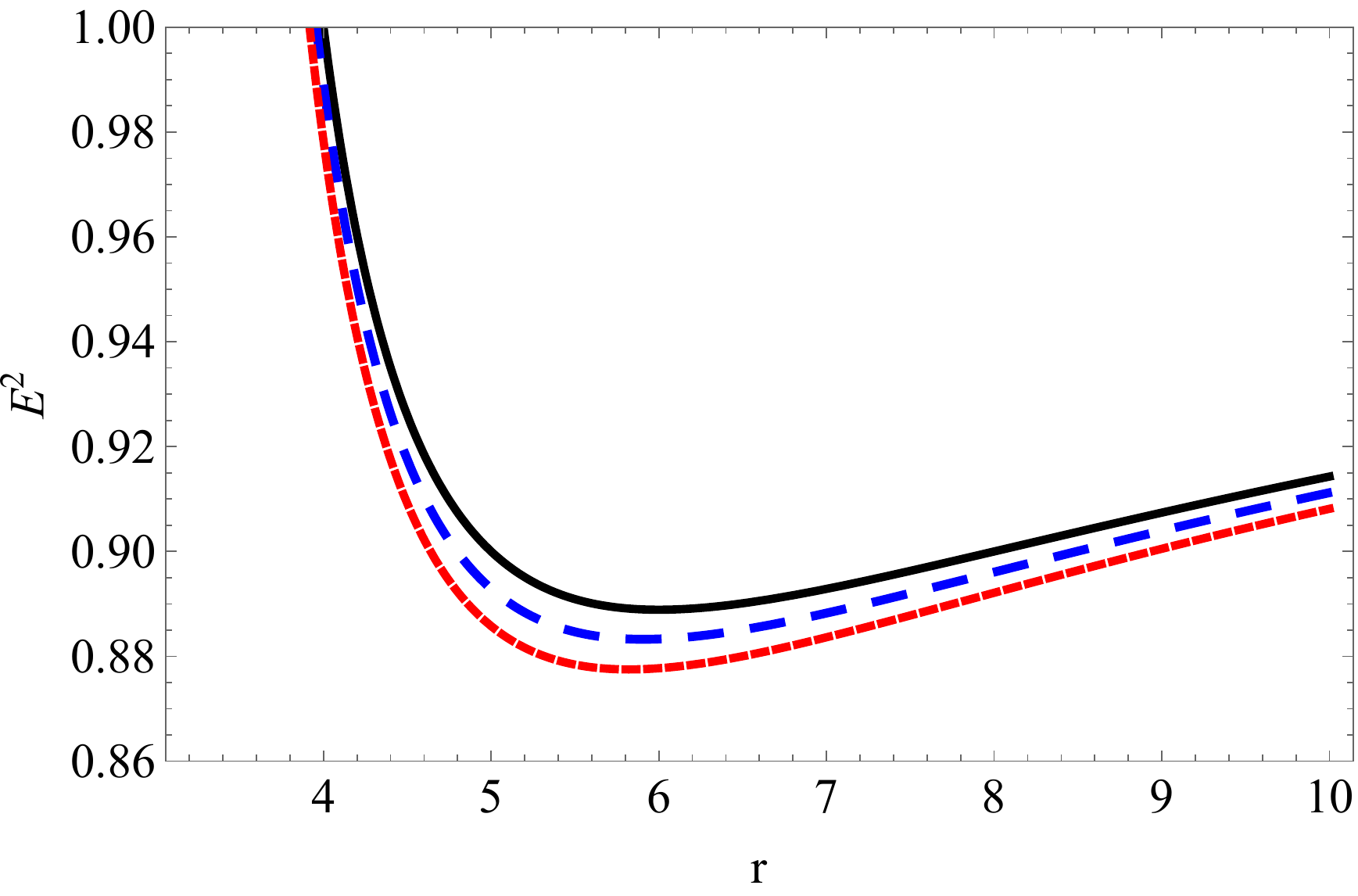}  \
\includegraphics[width=0.24\textwidth]{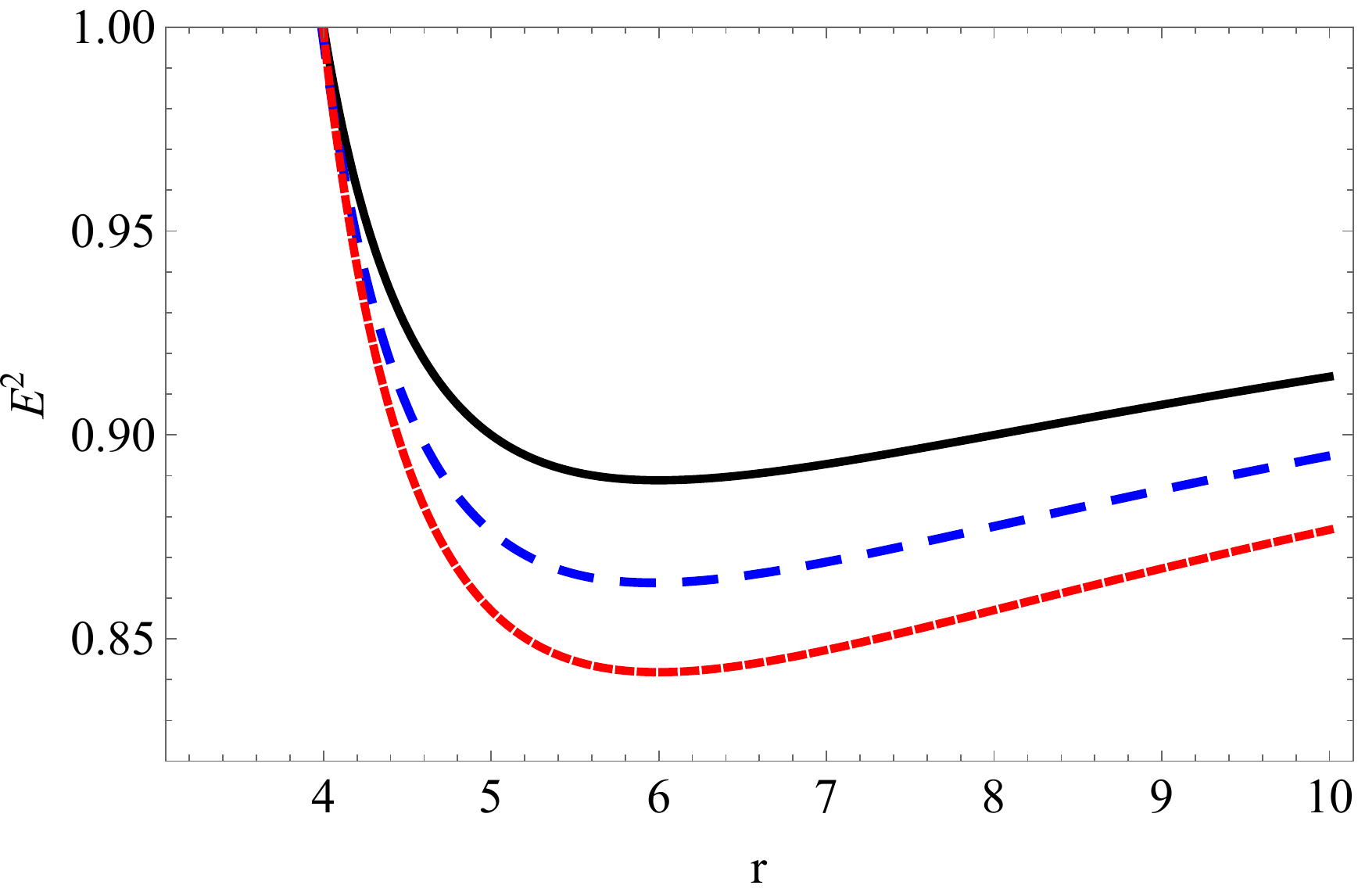}  \
\includegraphics[width=0.24\textwidth]{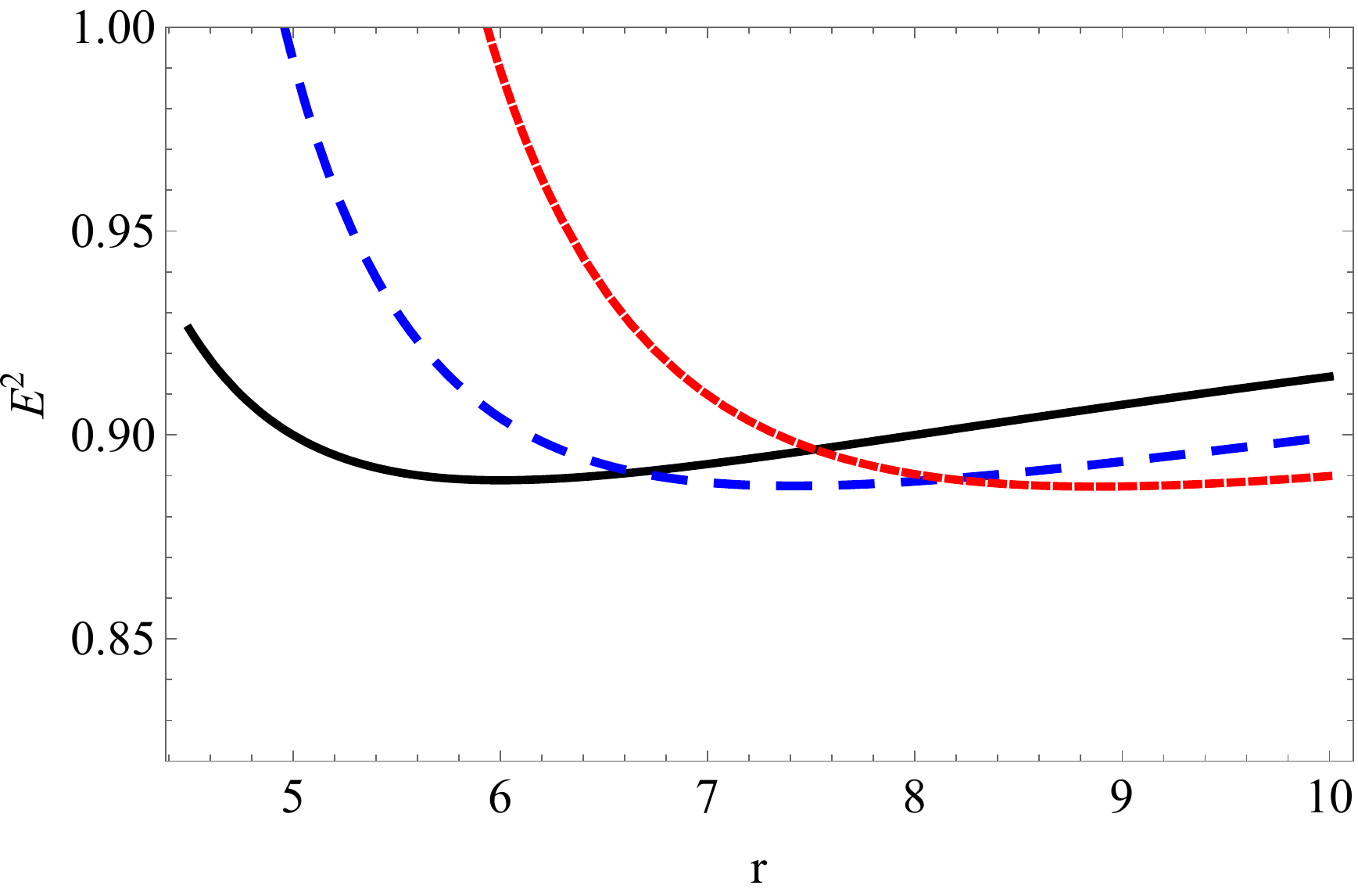} 
\caption{\label{LsqEsq} 
First row. $L^{2}$ for $M=1$, and $\alpha=0$ (black line, Schwarzschild), $\alpha=0.5$ (blue dashed line) and $\alpha=1$ (red dotted line) for model 1 (first panel) and model 2 (second panel) and 
and $\ell_{0}=0$ (black line, Schwarzschild),
$\ell_{0}=0.5$ (blue dashed line) and $\ell_{0}=1$ (red dotted line) for model 3 (third panel) and model 4 (fourth panel)\\
Second row. $E^{2}$ for $M=1$, and $\alpha=0$ (black line, Schwarzschild), $\alpha=0.5$ (blue dashed line) and $\alpha=1$ (red dotted line) for model 1 (first panel) and model 2 (second panel) and 
and $\ell_{0}=0$ (black line, Schwarzschild),
$\ell_{0}=0.5$ (blue dashed line) and $\ell_{0}=1$ (red dotted line) for model 3 (third panel) and model 4 (fourth panel)}. 
\end{figure*}

The profiles of the impact parameters for the Hairy BH's are illustrated in Fig. \ref{LsqEsq}. The first row corresponds to  $L^2$ parametrized with $\alpha$ (models 1 
and 2) and $\ell_0$ (models 3 and 4). The black line corresponds to the Schwarzschild solution. For models 1, 3 and 4 the location and value of the minimum of $L^2$ increase as $\{\alpha,\ell_{0}\}$ increase. Furthermore, in all the three cases, the minimum value that $L^{2}$ can reach corresponds to the black line (Schwarzschild case). In contrast, in model 2 the location and value of the minimum of $L^2$ decreases as $\alpha$ increases. Moreover, at $r \approx 6.5$ the three curves intersect and there is an interchange of the role played by the curves: black line becomes the lower bound for $r>6.5$.  The second row of Fig. \ref{LsqEsq} shows the profiles of $E^2$ for the four models with $M=1$. For models 1,2 and 3, the location of the minimum and value of $E^2$ decreases as either $\alpha$ or $\ell$ increases. However, model 4 exhibits an unconventional behavior as compared with the other three models. In this case it is shown a slight increase of the minimum of $E^2$ as $\ell_0$ increases and its location  shift significantly to the right.

\subsection{ISCO}
In this section, we discuss the behavior of the ISCO for timelike orbits and its relation with primary hairs $\{\alpha,\ell_{0}\}$. As discussed  previously, the ISCO corresponds to the last stable orbit, which corresponds to the inflection point of the effective potential, namely, $V_{eff}'' =0$ from Eq. (\ref{eq:veff}).
\begin{figure*}[hbt!]
\centering
\includegraphics[width=0.3\textwidth]{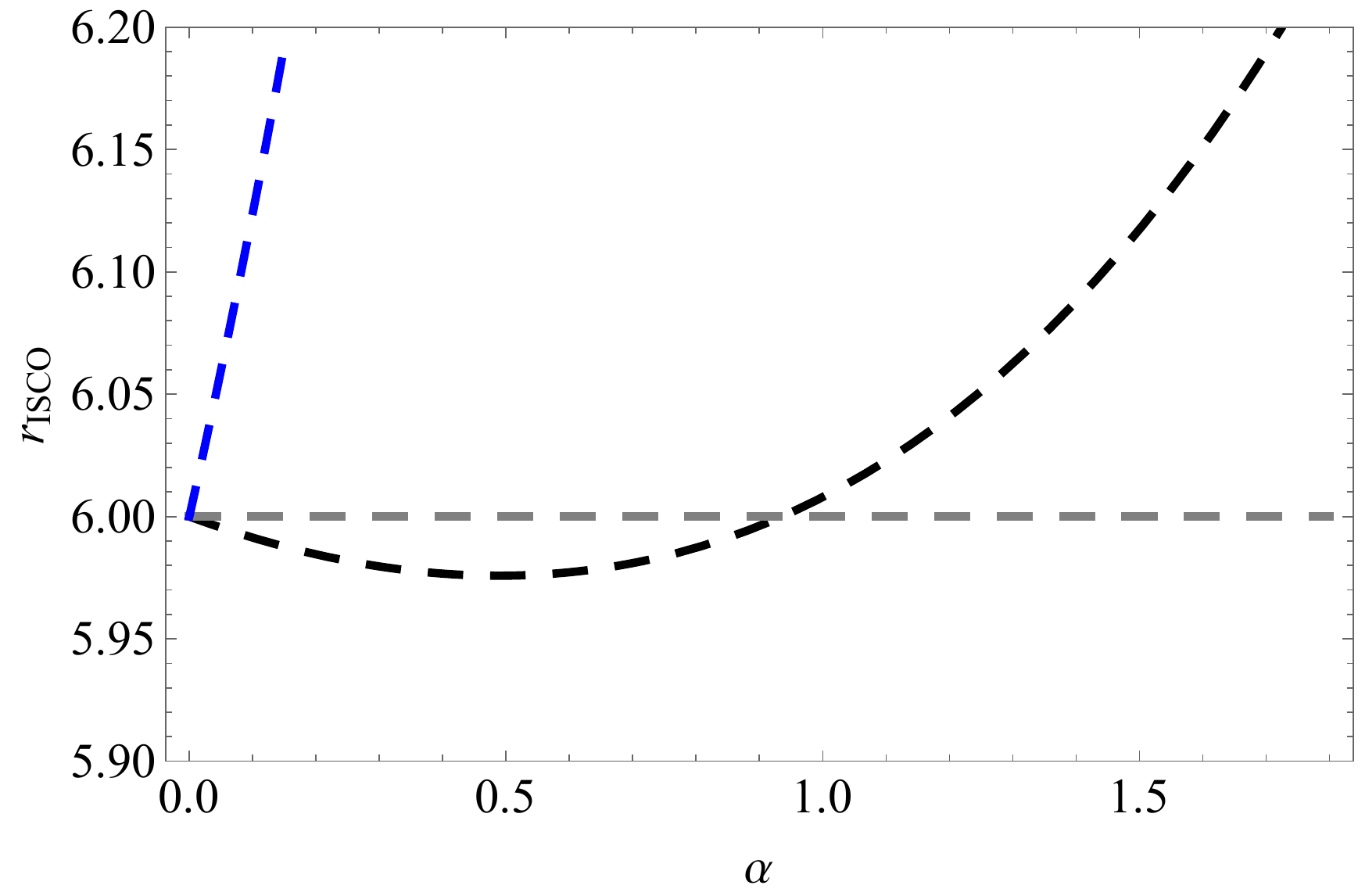}  \
\includegraphics[width=0.3\textwidth]{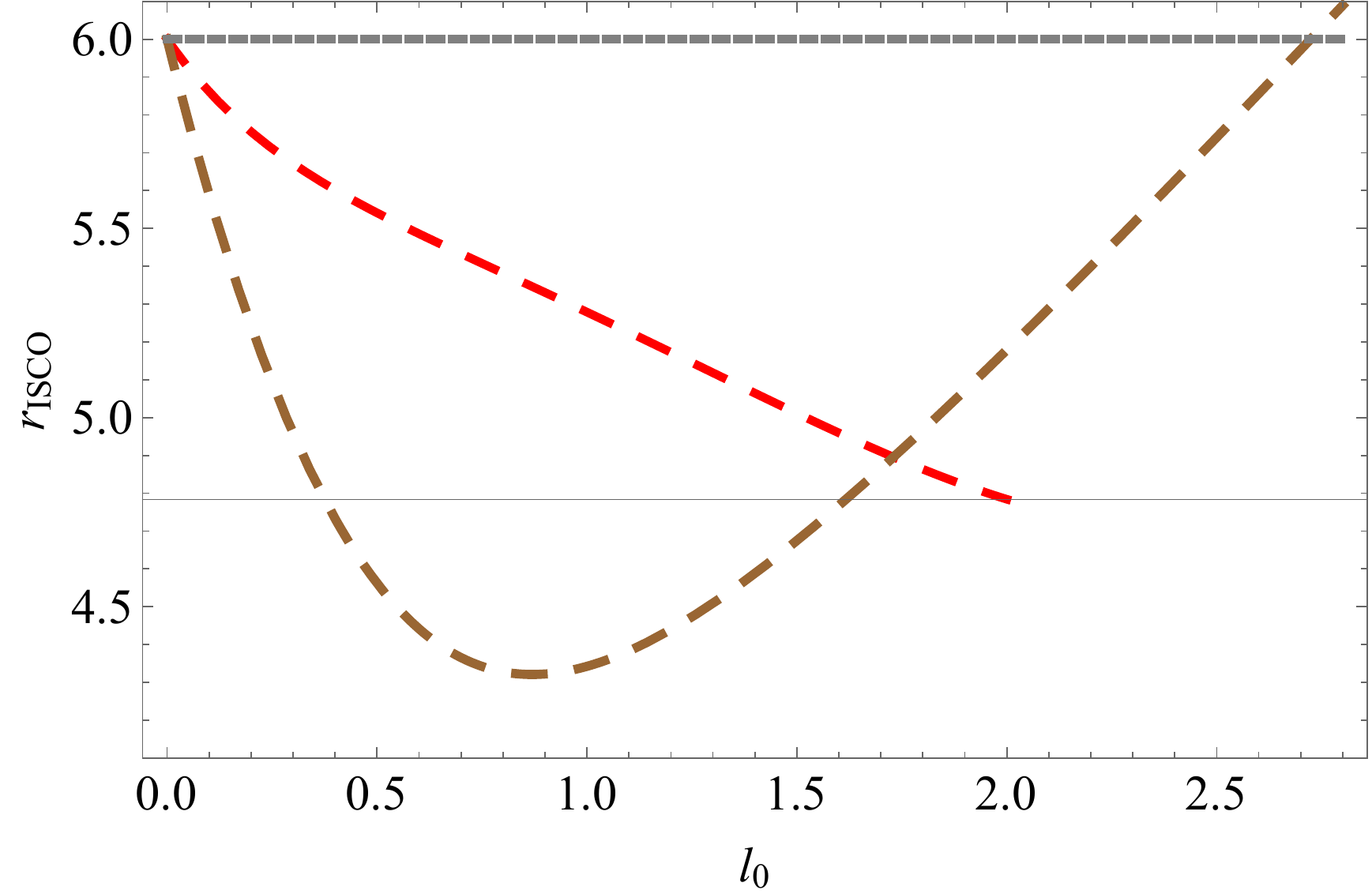}  \
\caption{\label{isco} 
ISCO radius for models 1 and 2 (black and blue lines in left panel, respectively) in dependence of the decoupling parameter $\alpha$ and for models models 3 and 4 (red and brown lines in right panel, respectively) in dependence of the decoupling parameter $\ell_{0}$
}. 
\end{figure*}
In Fig. \ref{isco} it is shown the ISCO radius, $r_{ISCO}$, for timelike orbits described by the particles around hairy BH's in the four models under consideration as a function of $\alpha$ (left panel, models 1 and 2) and $\ell_{0}$ (right panel, models 3 and 4). The gray dashed lines represent the ISCO for Schwarzschild solution located at $r_{ISCO}=6.0$. For model 1 (black dashed line), $r_{ISCO}$ decreases to some minimum value for some $\alpha$ and then behaves as a monotonously increasing function. In the second model (blue dashed line), $r_{ISCO}$ increases when $\alpha$ increases, but at some point it seems to diverge. For model 3 (red dashed line), the ISCO radius decreases as long as $\ell_0$ increases until it reaches 
the value of the Schwarzschild case (gray line). The last model (brown dashed line), describes a similar behaviour than model 1, namely $r_{ISCO}$ decreases until a minimum value for some $\ell_0$ and then behaves as a monotonously increasing function. The same behaviour has been reported in the case of timelike circular geodesics for charged particles around BH's geometries supported by a non--linear electrodynamics \cite{stuchlikprd}.

\subsection{MBO radius}
\begin{figure*}[hbt!]
\centering
\includegraphics[width=0.3\textwidth]{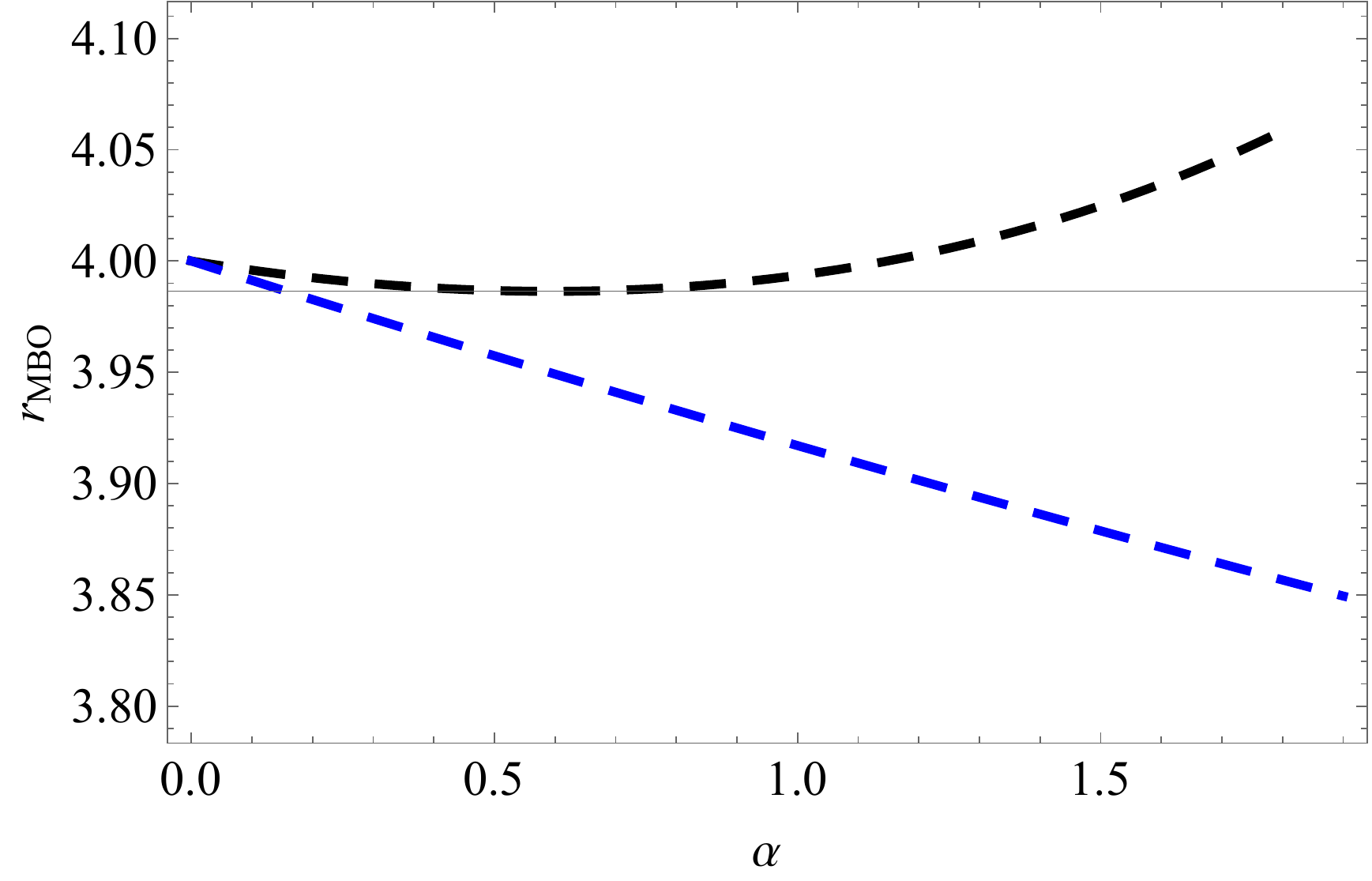}  \
\includegraphics[width=0.3\textwidth]{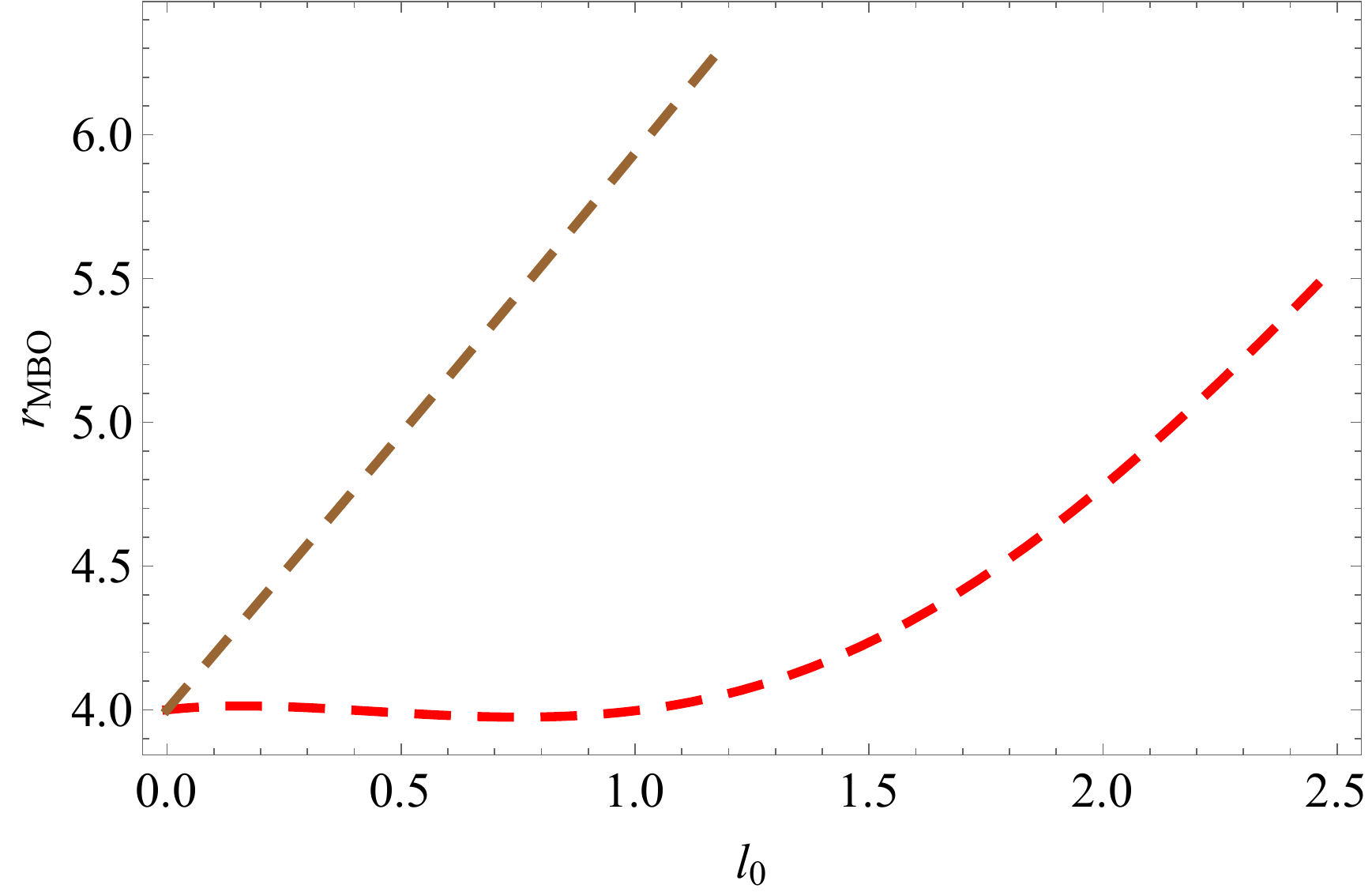}  \
\caption{\label{mb} 
MBO radius for models 1 and 2 (black and blue lines in left panel, respectively) in dependence of the decoupling parameter $\alpha$ and for models models 3 and 4 (red and brown lines in right panel, respectively) in dependence of the decoupling parameter $\ell_{0}$
}. 
\end{figure*}

In Fig. \ref{mb} we show the MBO radius, $r_{mbo}$, for the timelike orbits around the four different models under consideration. The gray dashed line corresponds to $r_{mbo}$ for the Schwarzschild case. In model 1 (black dashed line), $r_{mbo}$ decreases until it reach a minimum for some $\alpha$ (Schwarzschild line) and then it increases monotonously. In model 2 (blue dashed line), $r_{mbo}$ decreases monotonously as $\alpha$ increases until reach lower values than in the Schwarzschild case. In model 3 (red dashed line), $r_{mbo}$ is almost constant until some value of $\ell_0$ and then increases monotonously. In the last model (brown dashed line), $r_{mbo}$ increases as $\ell_0$ increases.

\section{Bounded orbits}\label{bo}

In this section
we analyze the bounded orbits for timelike geodesics around each hairy BH under consideration including the Schwarzschild solution. All the numerics have been performed by
considering $L = 4$ and $E \approx  0.95$.  

\begin{figure*}[hbt!]
\centering
\includegraphics[width=0.2\textwidth]{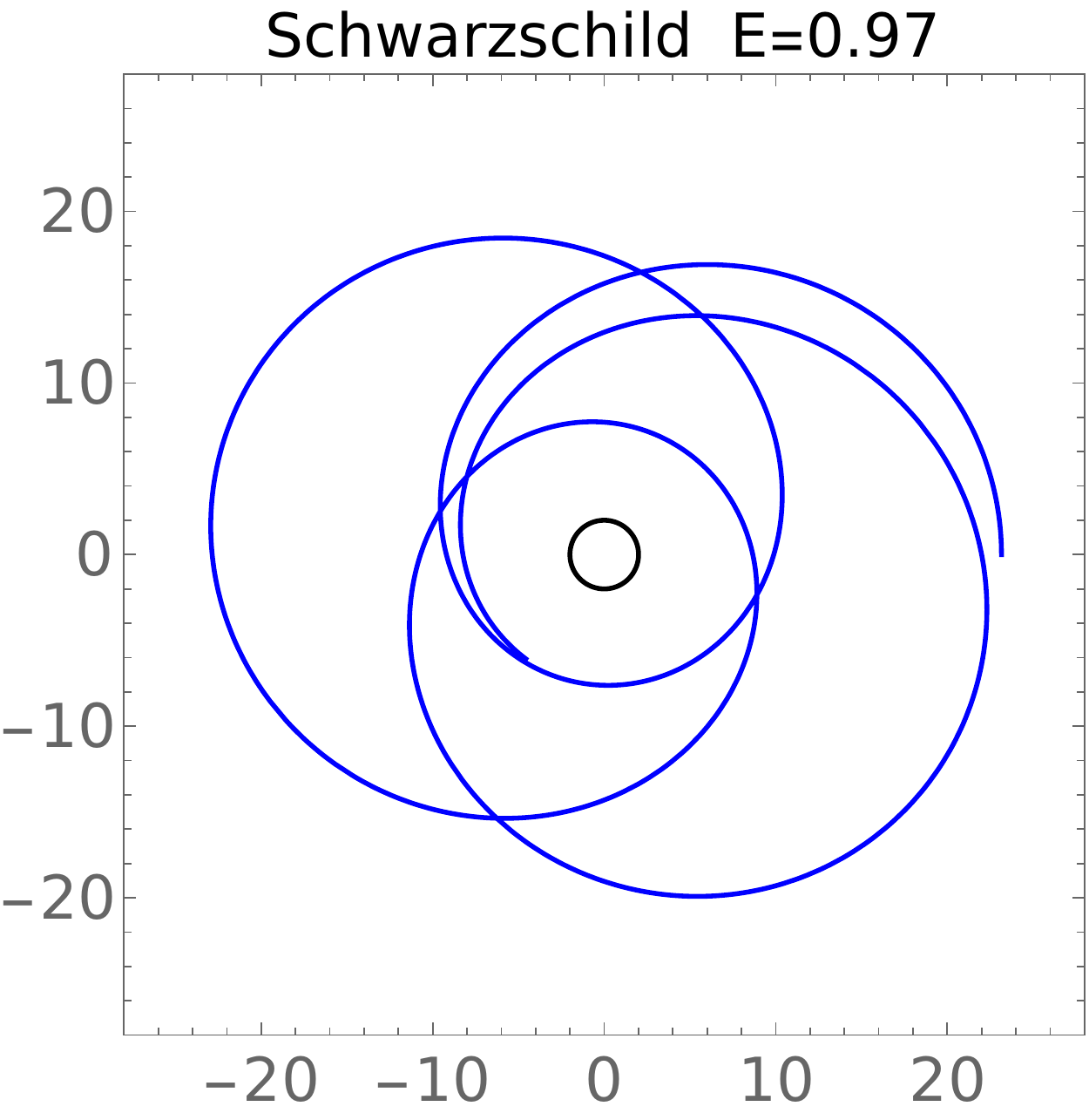}  \
\includegraphics[width=0.2\textwidth]{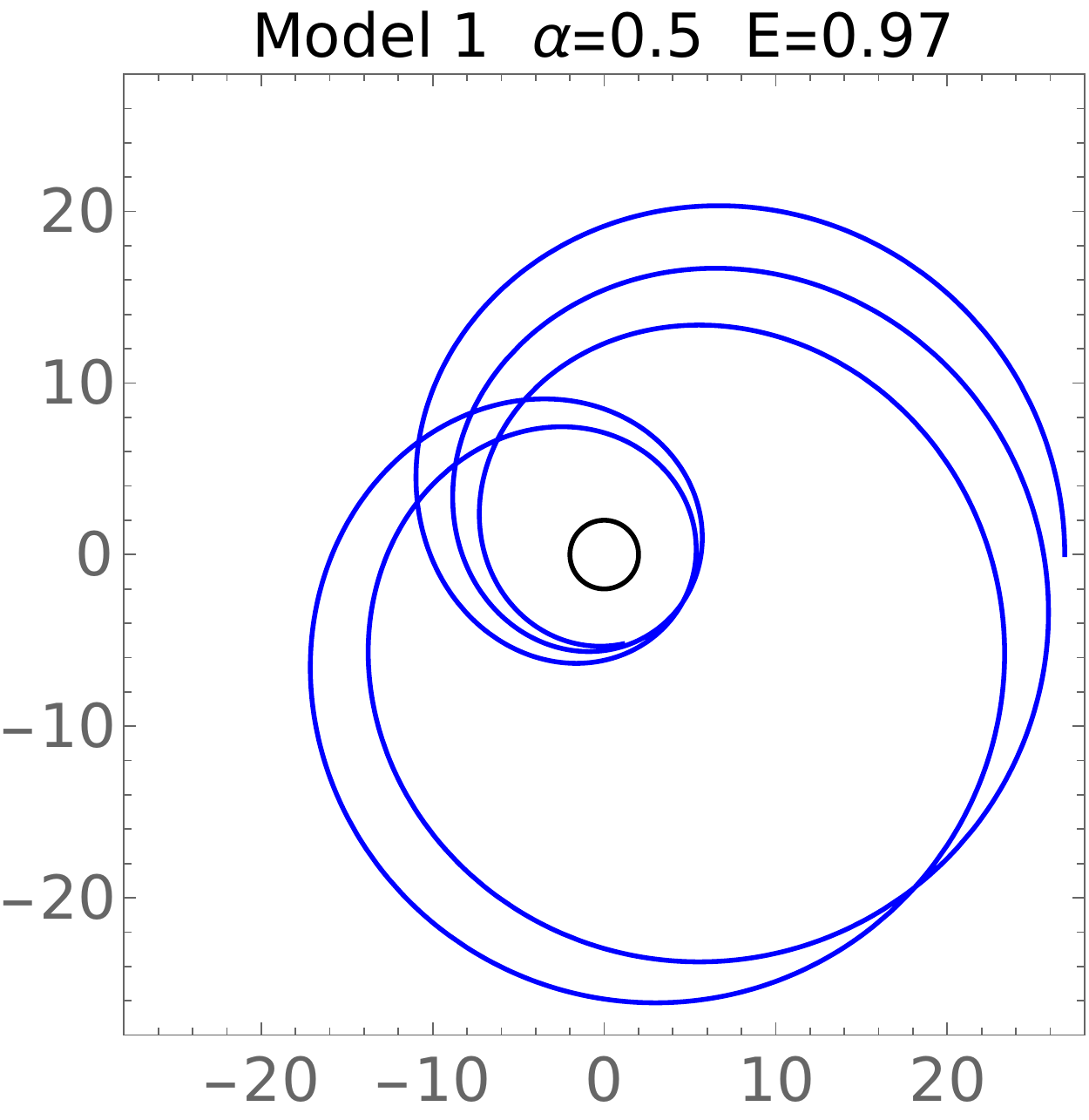}  \
\includegraphics[width=0.2\textwidth]{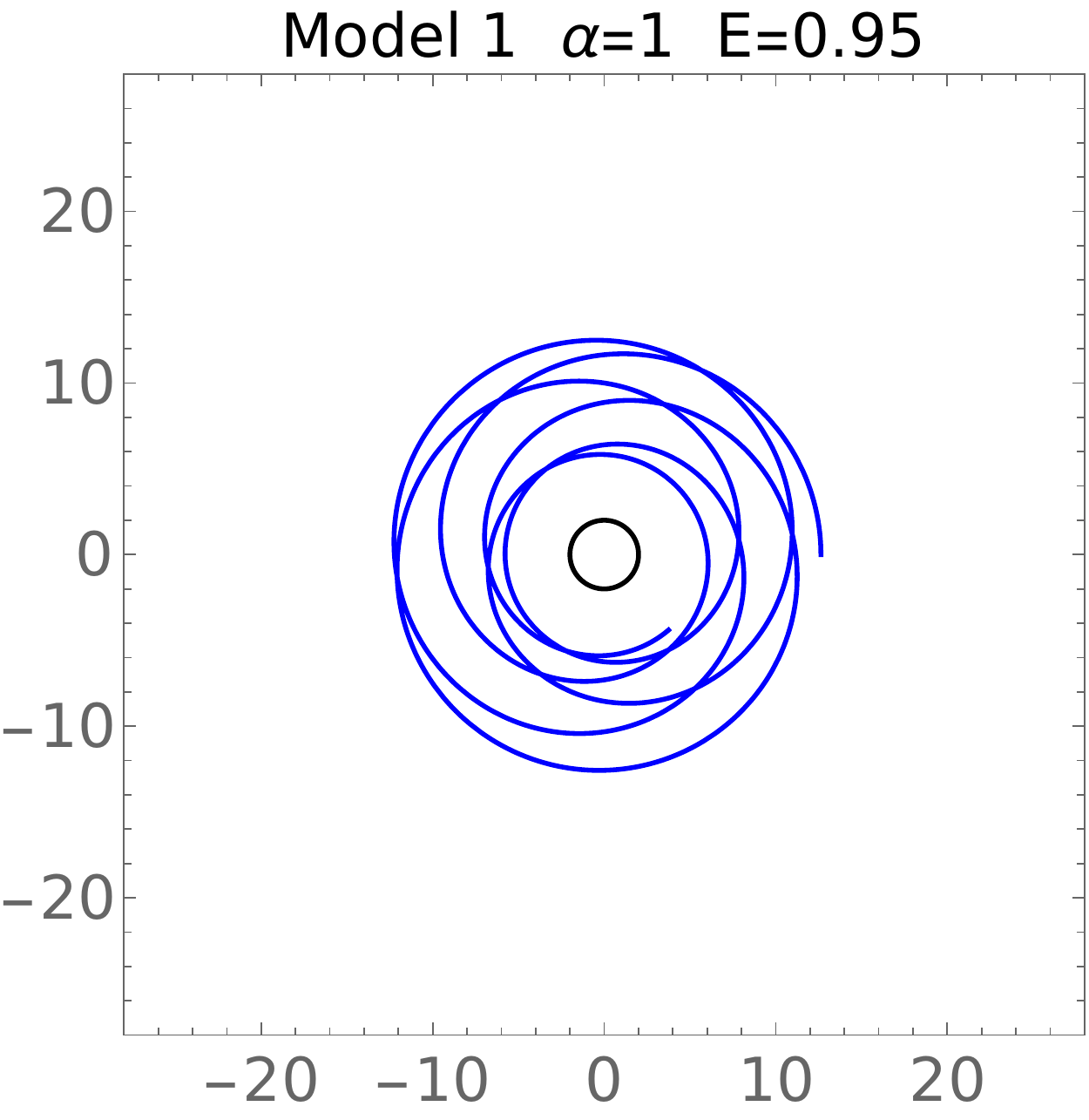}  \
\includegraphics[width=0.2\textwidth]{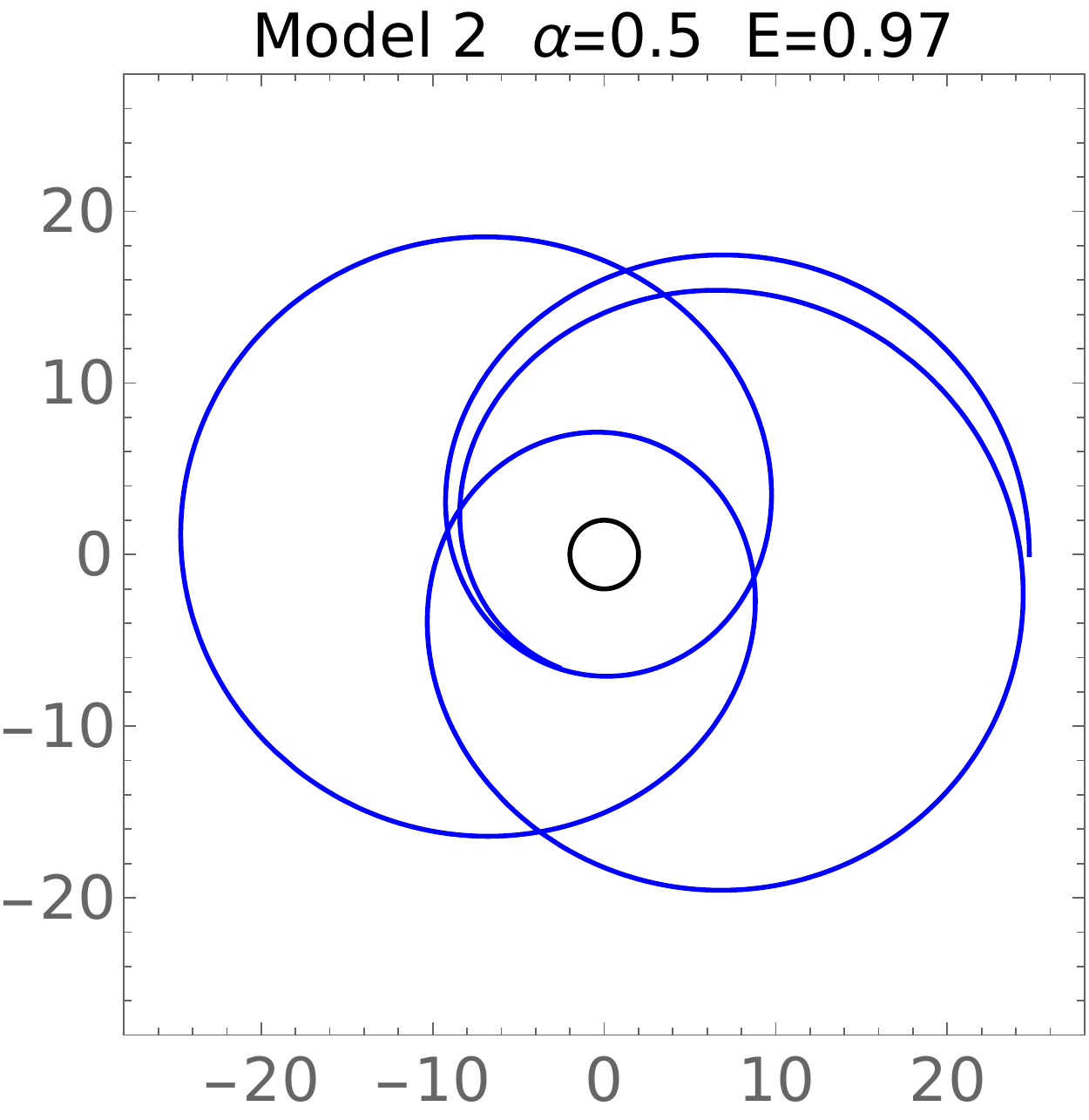}  \
\medskip

\includegraphics[width=0.2\textwidth]{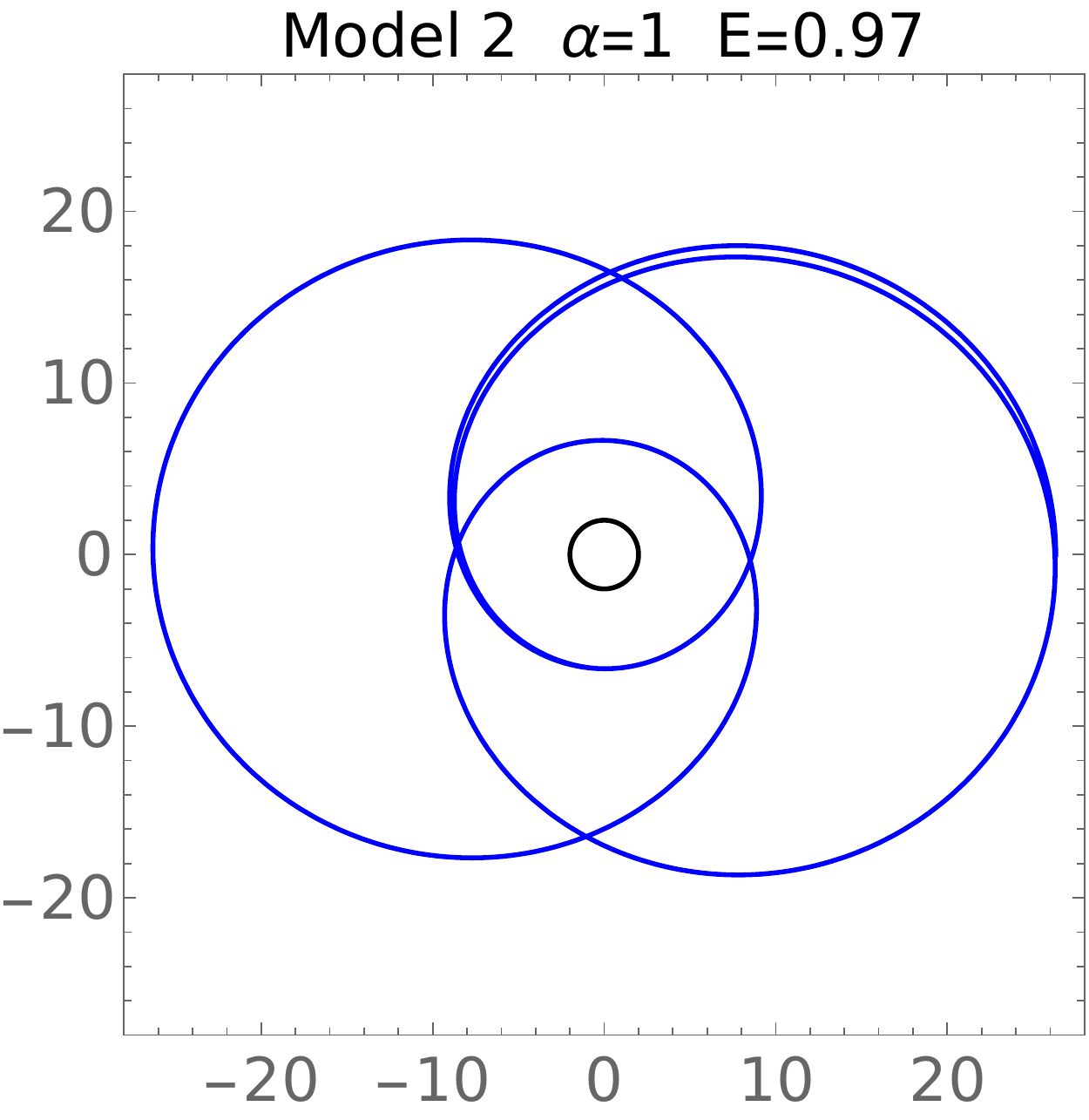}  \
\includegraphics[width=0.2\textwidth]{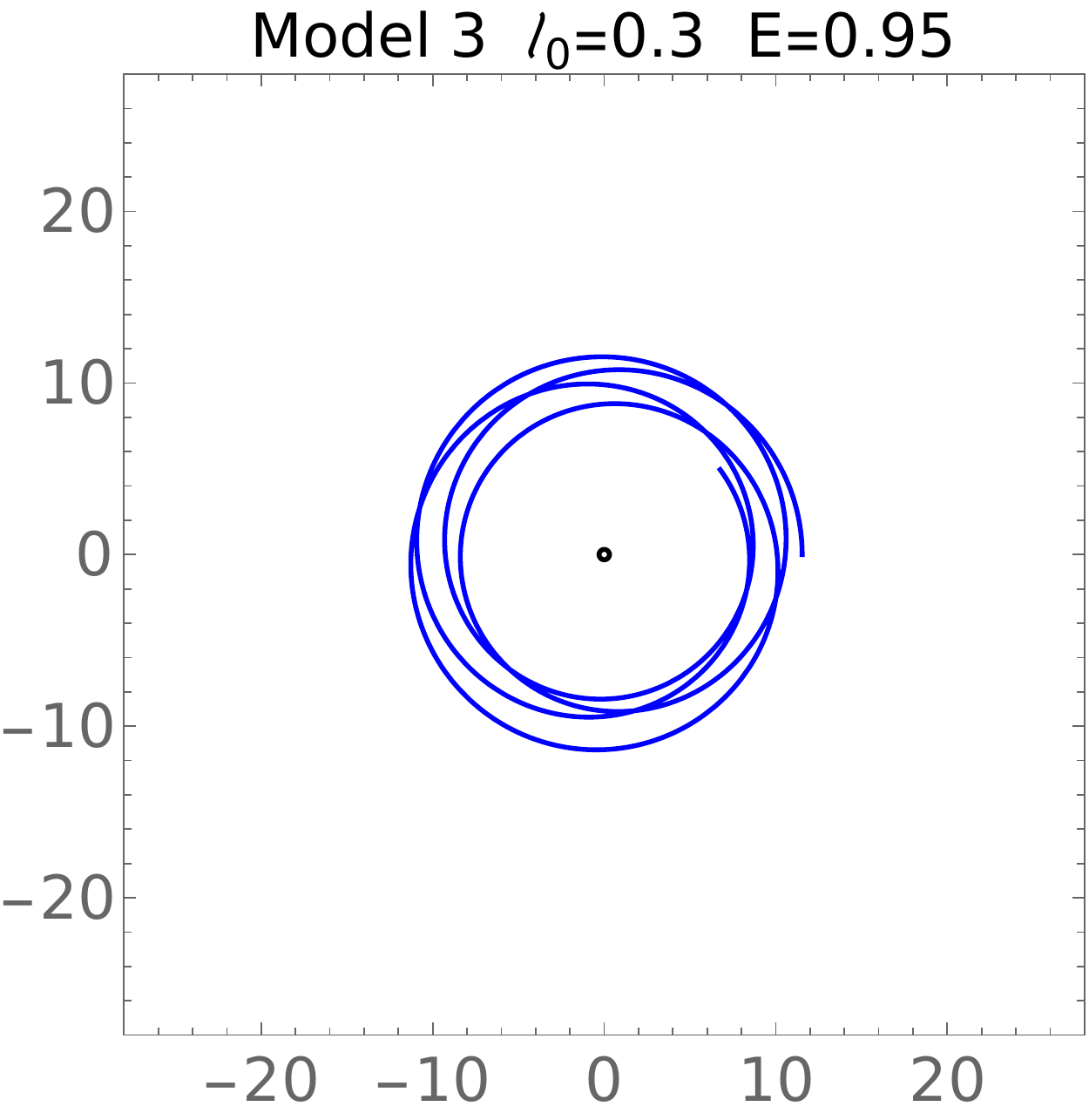}  \
\includegraphics[width=0.2\textwidth]{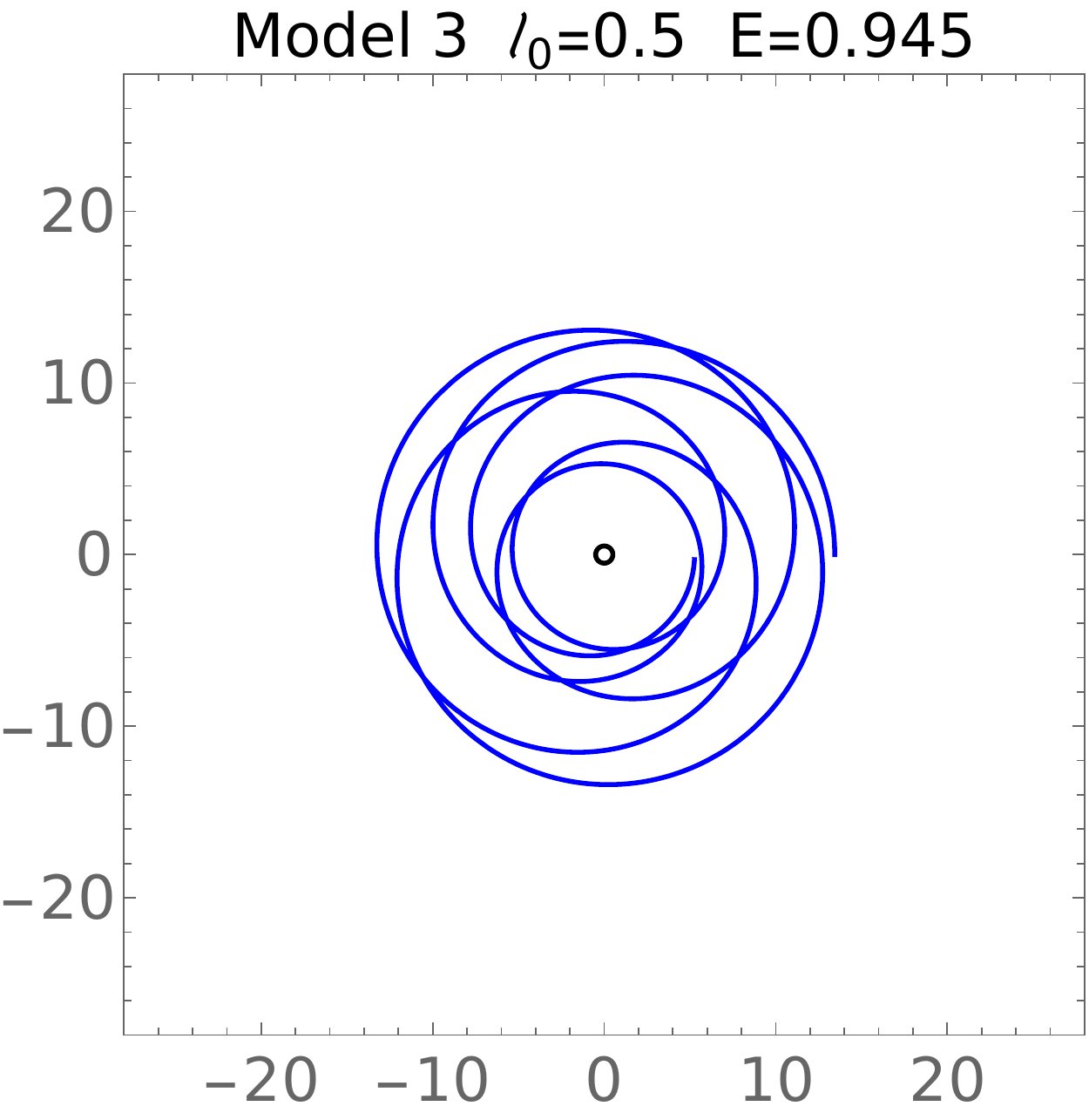}  \
\includegraphics[width=0.2\textwidth]{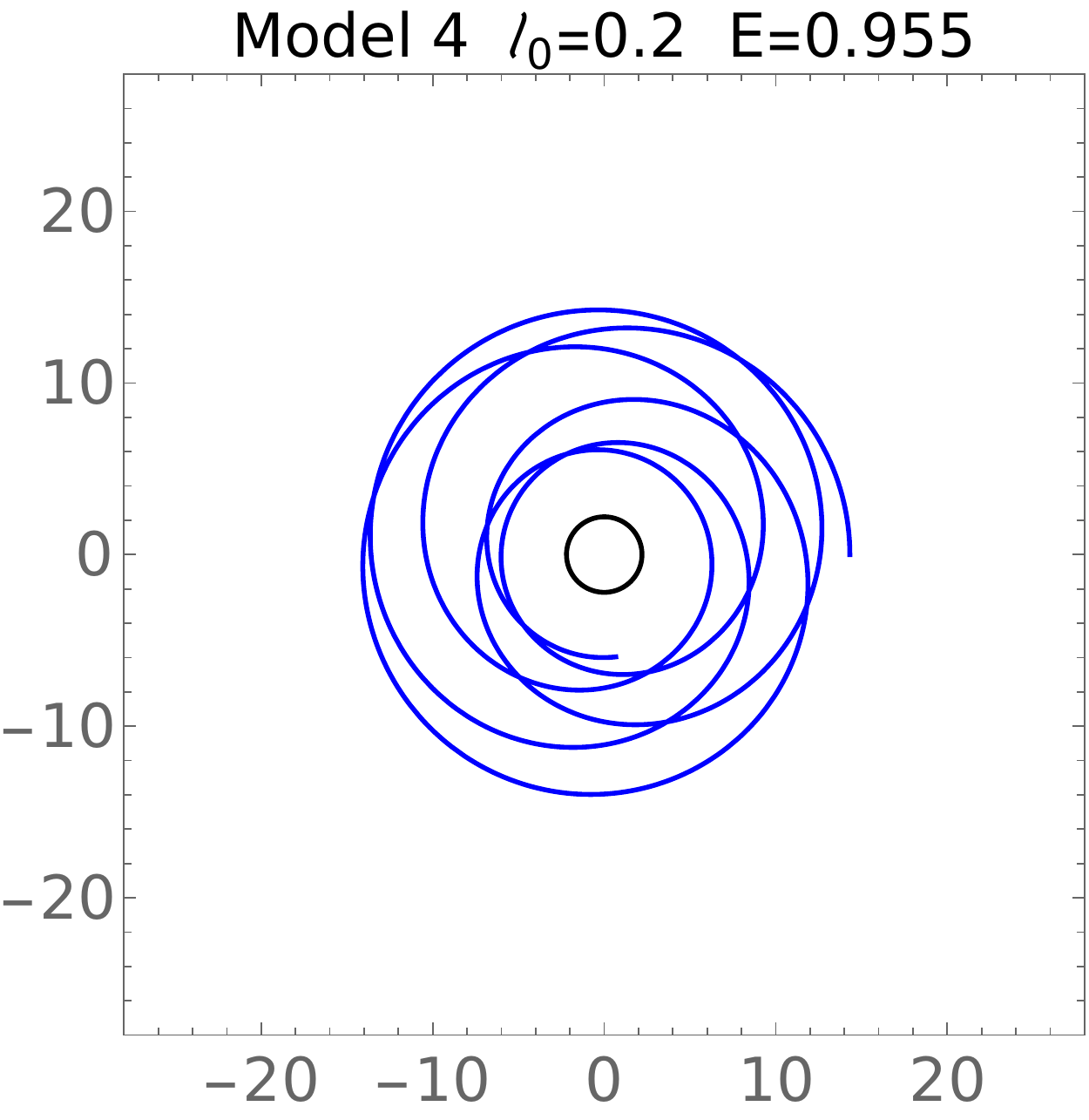} 
\caption{\label{BO}
Bounded orbits for different models, different energies $E$ and angular momentum of $L=4$. The first row shows a Schwarzschild BH, Model 1 and Model 2 BH. Second row shows Model 2, Model 3 and Model 4.}
\end{figure*}

As shown in figure \ref{BO}, all the orbits have an epicyclic motion with variation in the average size of the orbit in dependence of the parameters $\alpha$ and $\ell_0$. 
Model 1, in comparison with the Schwarzschild BH, shows an extension of the bound epicyclic motion to larger distances, which means that the particle-BH interaction becomes weaker. However, for increasing values of  $\alpha$, the size of the orbit gets reduced showing a strengthening of the particle-BH interaction. Model 2 shows a slight decrease in the bound of the epicyclic motion. Besides, the orbit shows an inverse relation between the average size radius and the parameter $\alpha$. In general, it is observed that this model represents a stronger particle-BH interaction in comparison with the Schwarzschild case. For model 3, both values of $\ell_0$ show a decrease in the average orbit radius with respect to the Schwarzschild case, suggesting a stronger interaction between the particle and the BH. Moreover, an inverse relation between $\ell_0$ and the average  radius of the orbit was observed. Finally, for $\ell_0 = 0.2$, model 4 presents a behaviour similar to model 3 for $\ell_0 = 0.5$, which means a stronger particle-BH interaction  in comparison with the Schwarzschild case. 

\section{Hairy black hole as a rotating solution mimicker}\label{mimic}
\begin{figure*}[hbt!]
\centering
\includegraphics[width=0.35\textwidth]{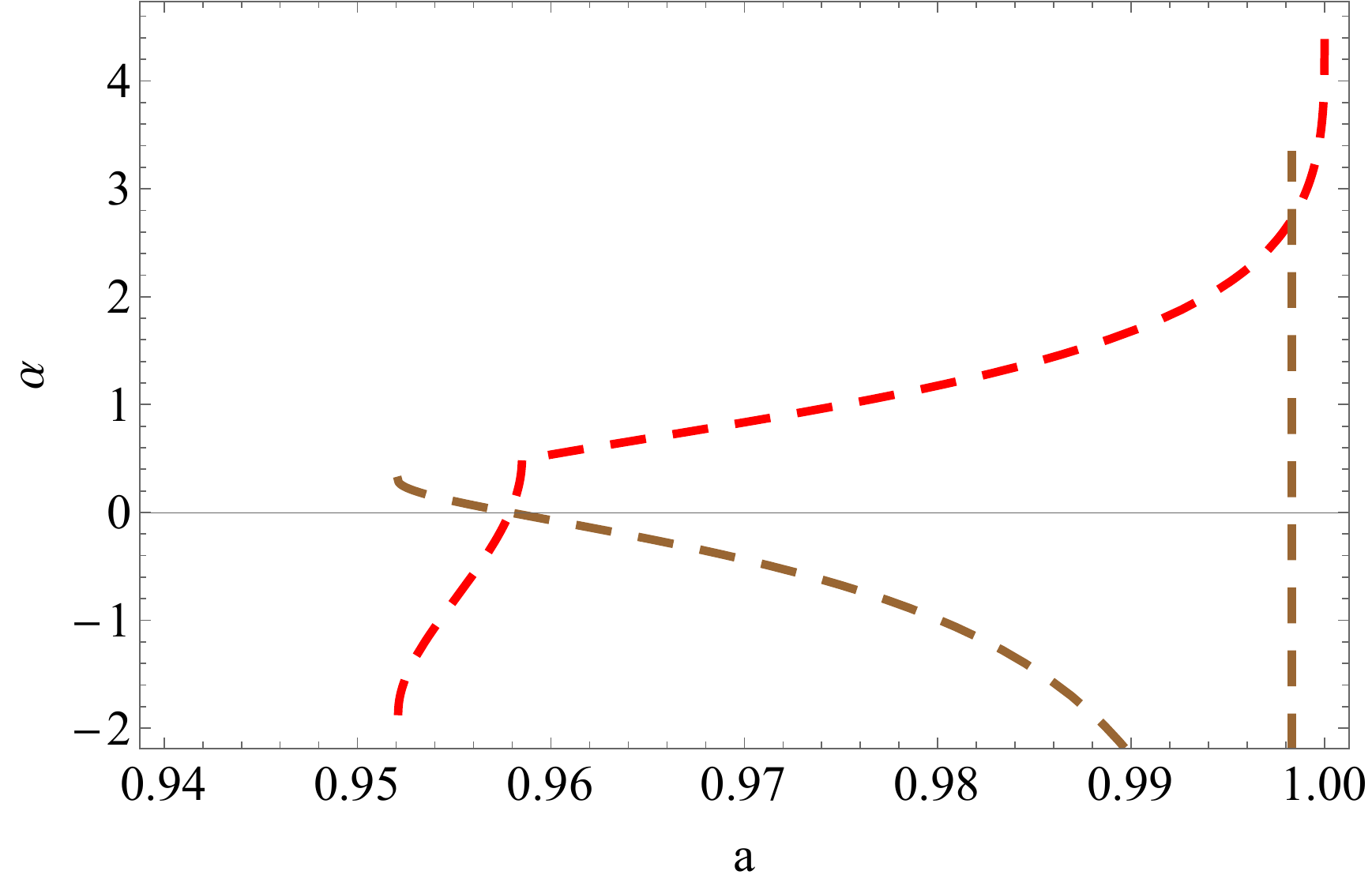}  \
\includegraphics[width=0.35\textwidth]{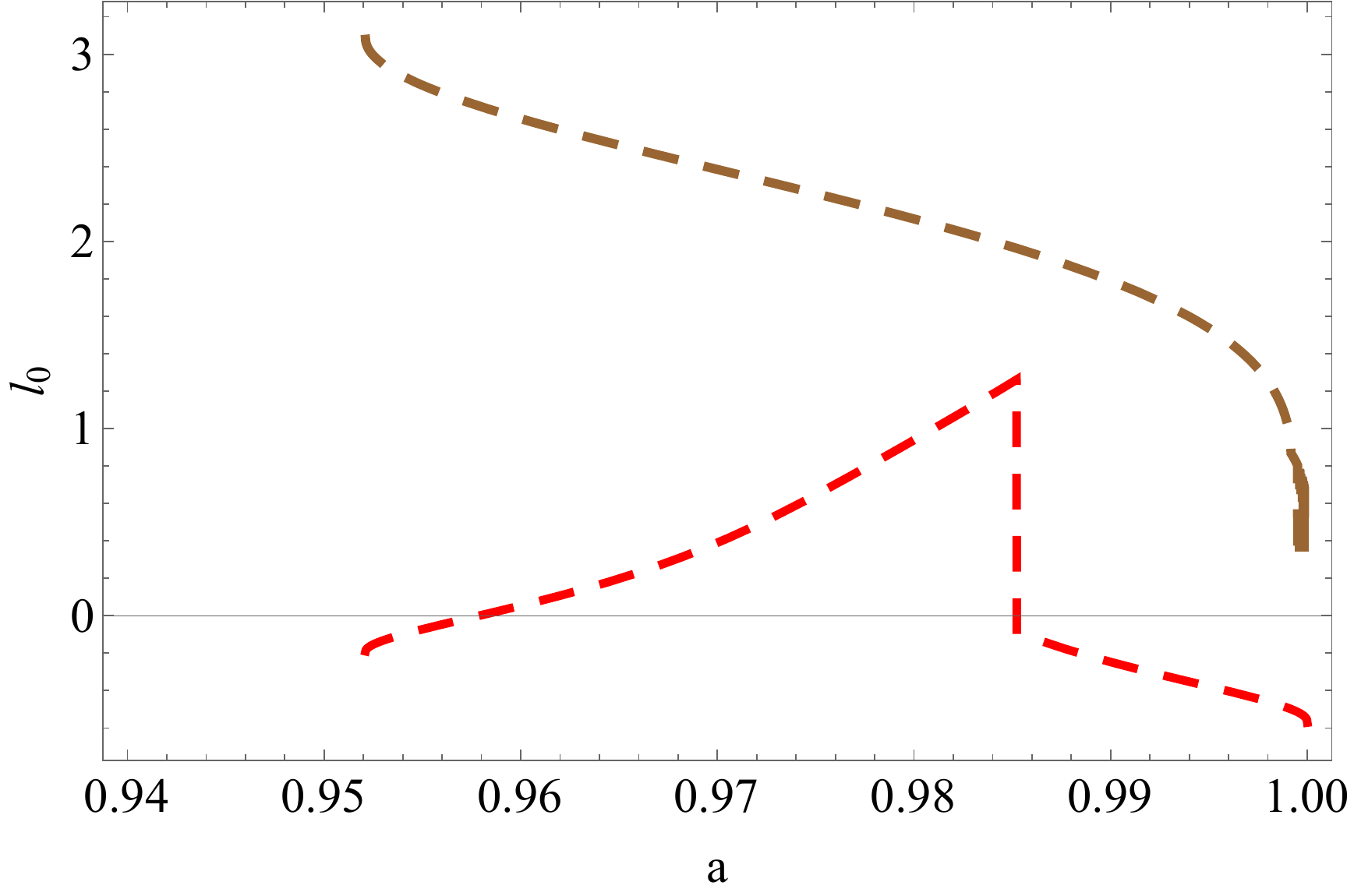}  \
\caption{\label{hair-kerr} 
The values of the primary hairs $\alpha$ (left panel) and $\ell_{0}$ (right panel) as a function of the rotation parameter $a$ giving the same ISCO radius. Red lines represent models 1 and 3. Brown lines corresponds to models 2 and 4.
}
\end{figure*}

The ISCO radius of test particles following corotating orbits around Kerr BH is given by
\begin{eqnarray}
r_{isco}=3+Z_{2}-\sqrt{(3-Z_{1})(3+Z_{1}+2Z_{2})},
\end{eqnarray}
where 
\begin{eqnarray}
Z_{1}&=&1+(\sqrt[3]{1-a}+\sqrt[3]{1+a})\sqrt[3]{1+a^{2}}\\
Z_{2}&=&\sqrt{3a^{2}+Z_{1}^{2}}
\end{eqnarray}
Now, it is possible to find a relationship between
the rotation parameter and the primary
hairs $\{\alpha,l_{0}\}$ such that the hairy BH found here serve as a mimicker of the rotating solution. In figure \ref{hair-kerr} we show the primary hairs for each model as a function of $a$.
In all the cases it is observed that the hairy BHs can mimic the rotation parameter in an interval 
$a\in(0.95,1)$ which corresponds to a high angular velocity regime. Interestingly, it is known that some active galactic nucleus as Ark564 and NGC1365 host a supermassive black hole with a high value of the spin parameter \cite{1,2,3,4}. For measurements of the spin parameter derived form relativistic reflection fitting of SMBH X-ray spectra
it have been obtained $a=0.96^{+0.01}_{-0.06}$ for Ark564 and $a=0.97^{+0.01}_{-0.04}$ for NGC1365 \cite{1,2,3,4}. In this work we use these values to estimate the primary hairs for rotating black hole mimickers associated to the models studied here (see table \ref{1}).  
\begin{center}
\begin{tabular}{cc|c|c|c|}
\cline{3-4}
& & $\alpha$ & $\ell_{0}$ \\ 
\cline{1-4}
\multicolumn{1}{ |c  }{\multirow{3}{*}{{\bf Ark 564}} }&
\multicolumn{1}{ |c| }{model 1} & 0.536154 & -      \\ 
\cline{2-4}
\multicolumn{1}{ |c  }{}                        &
\multicolumn{1}{ |c| }{model 2} & -0.0716532 & -       \\ 
\cline{2-4}
\multicolumn{1}{ |c  }{}                        &
\multicolumn{1}{ |c| }{model 3} & - & 0.0535898       \\ 
\cline{2-4}
\multicolumn{1}{ |c  }{}                        &
\multicolumn{1}{ |c| }{model 4} & - & 2.65561       \\ 
\cline{1-4}
\multicolumn{1}{ |c  }{\multirow{3}{*}{{\bf NGC 1365}} }&
\multicolumn{1}{ |c| }{model 1} & 0.835522 & -       \\ 
\cline{2-4}
\multicolumn{1}{ |c  }{}                        &
\multicolumn{1}{ |c| }{model 2} & -0.436882 & -       \\ 
\cline{2-4}
\multicolumn{1}{ |c  }{}                        &
\multicolumn{1}{ |c| }{model 3} & - & 0.389739       \\ 
\cline{2-4}
\multicolumn{1}{ |c  }{}                        &
\multicolumn{1}{ |c| }{model 4} & - & 2.38463       \\ 
\cline{1-4}
\end{tabular}
\captionof{table}{\label{1} Numerical mimickers values of the primary hairs for the four models considered.}
\end{center}
Note that, the hairy solutions in which we have based our study, depend on free parameters arising after GD is implemented, namely, the primary hairs $\alpha$ (for models 1 and 2) and $\ell_{0}$ (for models 3 and 4). In this respect, it is interesting how we can restrict these parameters to precise values
with observational data. Of course, we can use the same methodology to constraint the hairs by assuming other spin parameters. However, here we
have assumed Akr 564 and NGC 1365 to show that the procedure works.

\section{Null geodesics}\label{null}
\begin{figure*}[hbt!]
\centering
\includegraphics[width=0.3\textwidth]{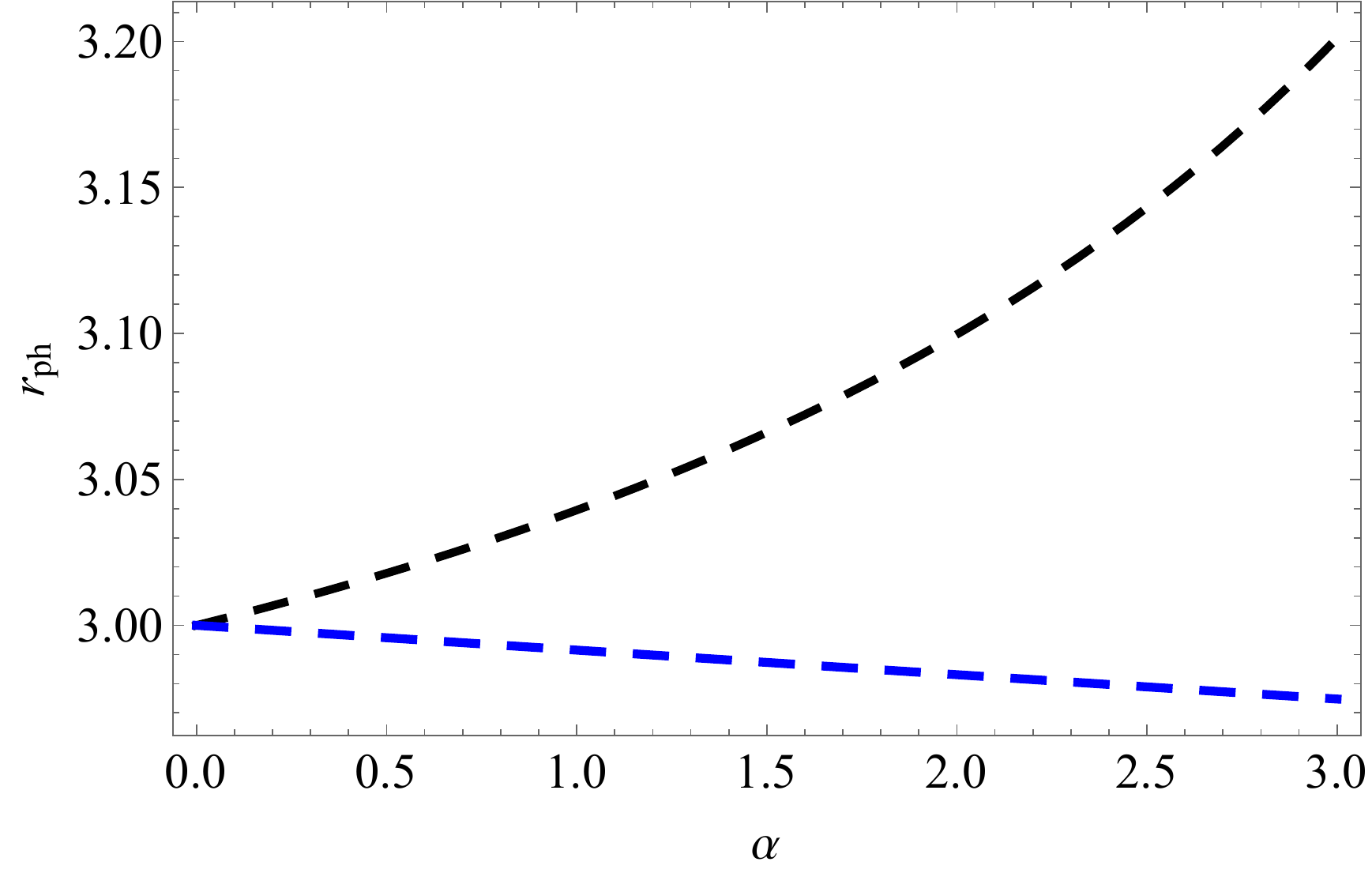}  \
\includegraphics[width=0.3\textwidth]{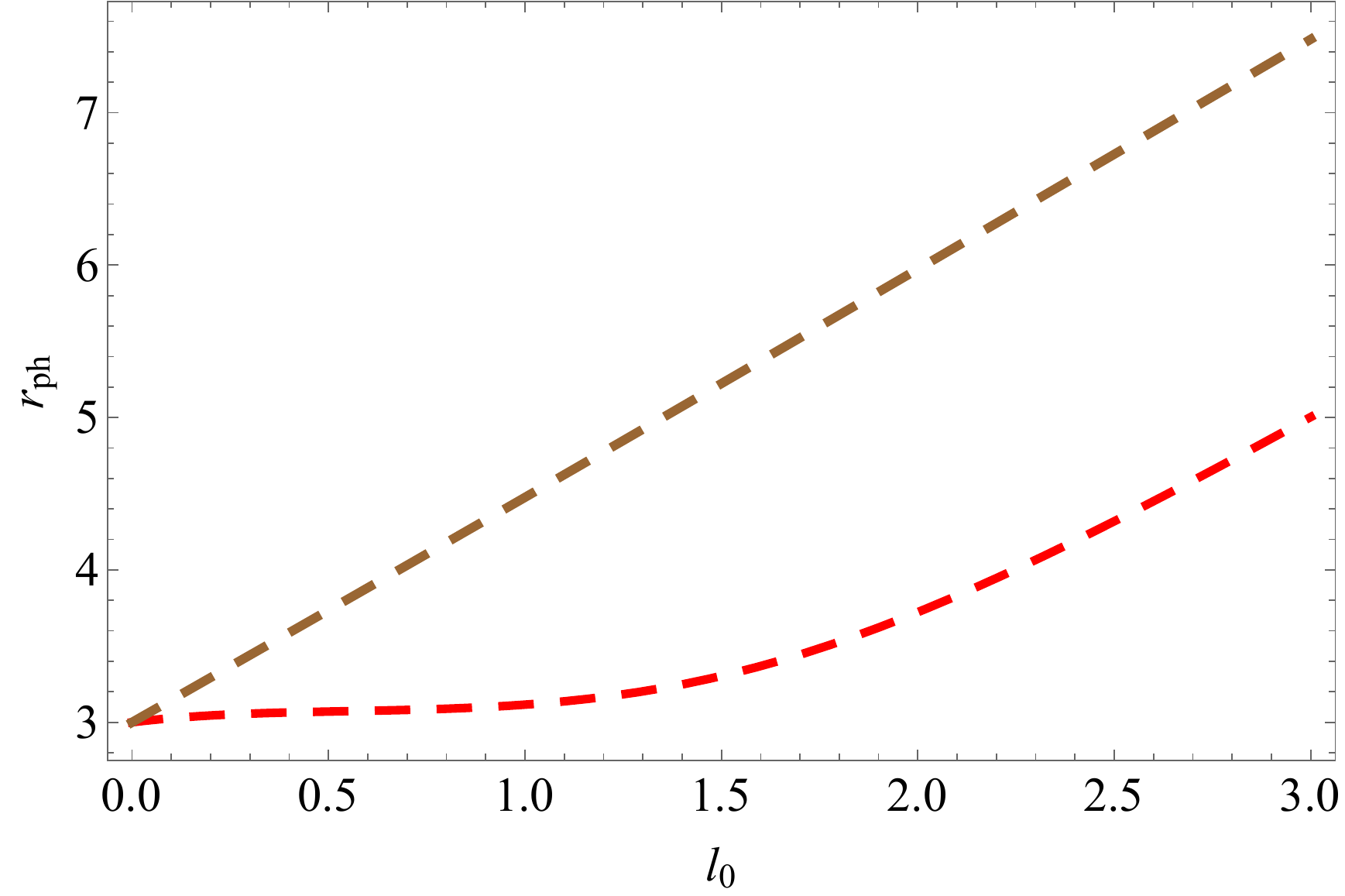}  \
\caption{\label{phs} 
Radius of the photon sphere. In left panel, black and blue line correspond to $r_{ph}$ for models 1 and 2 respectively in dependence of the decoupling parameter $\alpha$. In right panel, red and brown line correspond to $r_{ph}$ for models 2 and 3 respectively in dependence of $\ell_{0}$
}
\end{figure*}

\begin{figure*}[hbt!]
\centering
\includegraphics[width=0.3\textwidth]{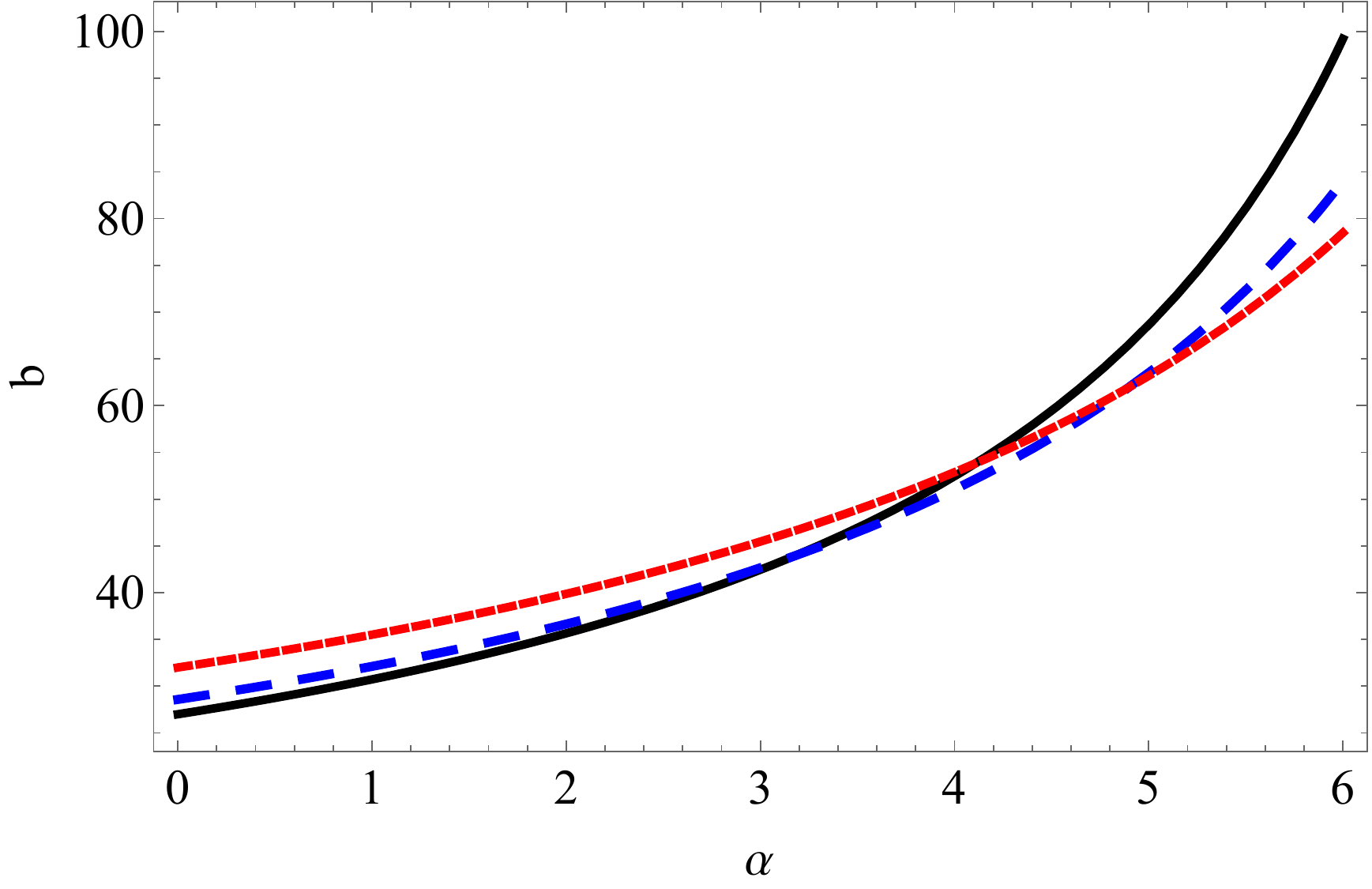}  \
\includegraphics[width=0.3\textwidth]{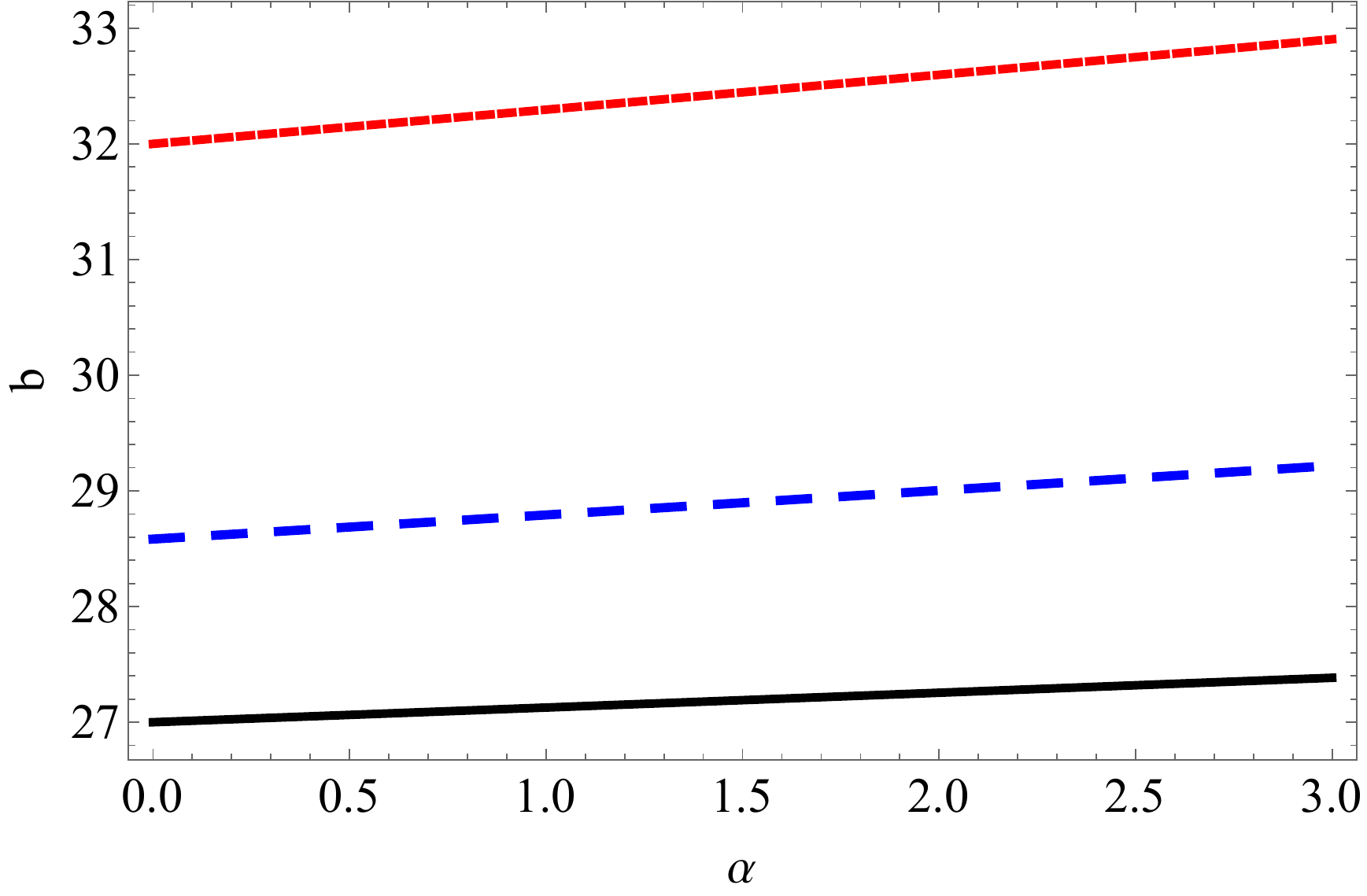}  \
\medskip

\includegraphics[width=0.3\textwidth]{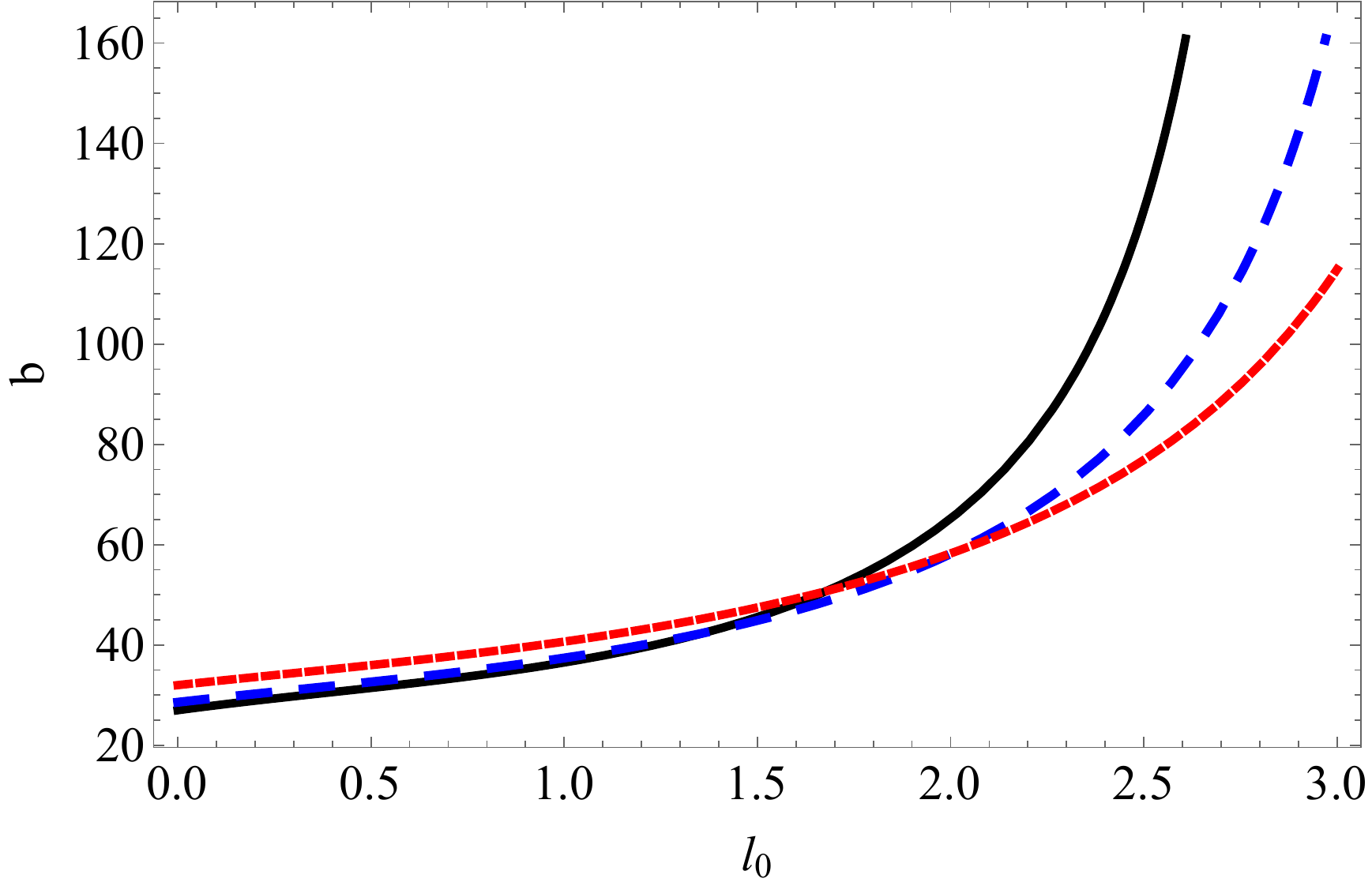}  \
\includegraphics[width=0.3\textwidth]{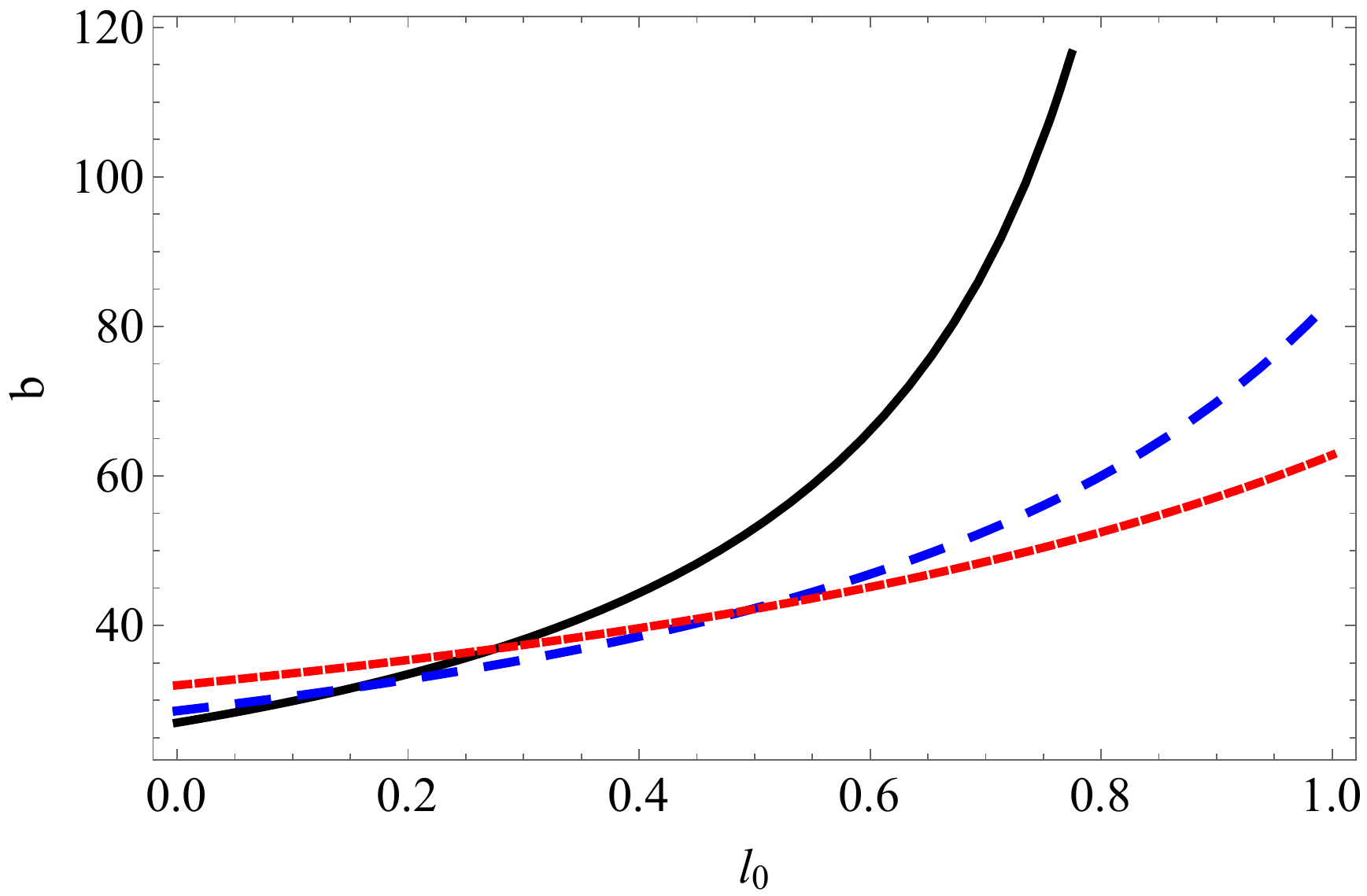} 
\caption{\label{b} 
Impact parameter for the different models considered. First row, left and right panel correspond to $b$ for models 1 and 2 respectively in dependence of $\alpha$. Second row, left and right panel correspond to $b$ for models 3 and 4 respectively in dependence of $\ell_{0}$
In all the plots, black, blue and red lines correspond to $r_{ph}=3$, $r_{ph}=3.5$ and $r_{ph}=4$ respectively.
}
\end{figure*}

In this section we study the null geodesics in order to get the photon radius $r_{ph}$ for the different models here considered. The null geodesic equation is obtained when vanishing $\epsilon$, $\mathcal{Q}$ and $\dot{r}$ in Eq. \ref{rdot}, which becomes
\begin{equation}
    r=\sqrt{f}\frac{L}{E}. \label{nullgeo}
\end{equation}
In figure \ref{phs} it is shown the radius of the photon sphere corresponding to each of the four models considered. We have as reference the well known radius of the Schwarzschild photon sphere which corresponds to $r_{ph} = 3$. On one hand, in the left panel the $r_{ph}$ for model 1 and model 2 are represented as a function of the parameter $\alpha$. It can be appreciated that for model 1 (black dashed line) the photon sphere radius increases with $\alpha$, meanwhile for model 2 (blue dashed line) this radius decreases. On the other hand, in the right panel the corresponding $r_{ph}$ for model 3 and model 4 are represented in dependence of the parameter $\ell_{0}$. In this case, we observe that for both model 3 (red dashed line) and model 4 (brown dashed line), the photon sphere radius increases as long as $\ell_{0}$ increases. However, model 4 increases faster than model 3.
In figure \ref{b} we show the impact parameter $b$ considering three values of $r_{ph}$ for each of the four models studied. 
The impact parameter is important because it governs the size of the shadow of the corresponding BH and it is given by
\begin{equation}
    b=\frac{r^2}{f},
\end{equation}
where $b=L/E$.  In the first row we plot the parameter $b$ in dependence of $\alpha$, for model 1 and model 2 respectively. In model 1 we observe that the impact parameter $b$ always increases with $\alpha$. Moreover, considering different values of photon sphere radius, when $r_{ph}>3$ (Schwarzschild case) we get higher values of $b$ only when we consider $0<\alpha\leqslant 3$; for bigger values of $\{\alpha,b\}$, $r_{ph}$ is smaller. In the same way, in model 2 we see that as long as $\alpha$ increases, the parameter $b$ slightly increases. Moreover, there is a drastic change in $b$ when the photon sphere radius is increased, so higher values of $r_{ph}$ lead higher impact parameters. In the bottom panels we plot the parameter $b$ in dependence of $\ell_{0}$, for model 3 and model 4 respectively. In model 3 we observe that the impact parameter $b$ increases with $\ell_{0}$. Besides, if we plot it for different values of the photon sphere radius, when $r_{ph}>3$ (Schwarzschild $r_{ph}$) we get higher values of $b$ only when we consider $0<\ell_{0}\leqslant 1$. For bigger values of $\ell_{0}$ the behavior is reverted which means that $b$ increases faster for smaller $r_{ph}$ values. A similar behavior describes the last model in which it is observed that as long as $\ell_{0}$ increases, the parameter $b$ increases as well. Moreover, considering different values of $r_{ph}$, we see that when the photon sphere radius is increased, the impact parameter $b$ also increases but only in the $0<\ell_{0}\leqslant 0.1$ domain. For higher values of $\ell_{0}$ the behavior is reverted and we can appreciate that smaller values of $r_{ph}$ lead higher impact parameters $b$.

\section{Conclusions}\label{concl}
Is well-known 
 that the apparition of stable matter/field configurations around the
black hole geometry is possible, leading thus to the so–called
hairy black holes solution. This was precisely the topic
under study in this paper. In this work we considered hairy black holes which correspond to deviation from the Schwarzschild geometry by a generic fluid which can be though as a kind of non--linear electrodynamics. Besides, we performed a complete analysis regarding to
the behaviour of both time--like and null geodesics and obtained that the behaviour of test particles are appreciably different when compared with the path they follow around a Schwarzschild black hole. All of these results lead to conclude that the appearance of primary hair $\{\alpha,\ell_{0}\}$ by gravitational decoupling plays a critical role in the behaviour of test particles around black holes.

In order to restrict the values of the primary hairs of the solution we considered their relation with the rotating parameter of the Kerr black hole giving the same innermost stable circular orbit radius. As a result we obtained that such a relation is possible whenever the spin parameter belong to the interval $a\in(0.95,1)$ which corresponds to a high rotating velocity regime. Besides we found the numerical values for the hairs $\{\alpha,\ell_{0}\}$ using the data of the supermassive black holes in the active galactic nucleus Ark 564 and NGC 1365.

Besides, it could be interesting to explore if for some values of the parameters $\{\alpha,\ell_{0}\}$, the hairy solutions can mimic a rotating geometry as the Kerr BH, for example.

Before concluding this work, we would like to recall that the hairy black hole here considered can be interpreted as solutions supported by some non--linear electrodynamics to some extent. In this sense, the analysis of null geodesics followed by photons should be modified in order to take into account the non--linear effects  as done in Ref. \cite{stuchlikprd}. 

However, the discussion of these and other aspects go beyond the scope of this paper and we leave this discussion for a future work.



\end{document}